\RequirePackage{rotating}

\documentclass{svjour3}
\usepackage[table]{xcolor}

\usepackage{amssymb}
\usepackage{xcolor}
\usepackage{booktabs}
\usepackage{lscape}
\usepackage{url} 
\usepackage[htt]{hyphenat}
\usepackage{amsmath}
\usepackage[misc,geometry]{ifsym}
\usepackage{tcolorbox}
\usepackage{rotating}
\usepackage{float}
\usepackage{placeins}
\usepackage{longtable}
\usepackage{chngcntr}
\usepackage[justification=centering]{caption}
\usepackage[shortlabels]{enumitem}
\usepackage{pdflscape} 
\usepackage{adjustbox}
\usepackage[utf8]{inputenc}

\usepackage{multirow}
\newtcolorbox{mytextbox}[1][]{%
  sharp corners,
  colback=white,
  attach title to upper,
  #1,
  fontupper=\footnotesize
}

\newcommand{\Ahmed}[1]{\textcolor{brown}{{\it [Ahmed: #1]}}}

\begin{document}

\title{Exploring Data Management Challenges and Solutions in Agile Software Development: A Literature Review and Practitioner Survey}
\titlerunning{ Data Management Challenges and Solutions in Agile Software Development}

\author{Ahmed Fawzy    \and
        Amjed Tahir     \and
        Matthias Galster    \and
        Peng Liang  
}

\institute{Ahmed Fawzy \and Amjed Tahir  \at School of Mathematical and Computational Sciences \\  Massey University, New Zealand \\ 
            \email{ahmed.mohamed.6@uni.massey.ac.nz, a.tahir@massey.ac.nz}
            \and
           Matthias Galster \at Department of Computer Science and Software Engineering \\ University of Canterbury, New Zealand \\ 
            \email{mgalster@ieee.org}
            \and
           Peng Liang \at School of Computer Science \\ Wuhan University, China \\ 
            \email{liangp@whu.edu.cn}
}

\date{Received: date / Accepted: date}

\maketitle
\begin{abstract}

$\ $
\begingroup 
\color{black}\\ 
\noindent \textit{\textbf{Context: }}
Managing data related to a software product and its development poses significant challenges for software projects and agile development teams. These include integrating data from diverse sources and ensuring data quality amidst continuous change and adaptation.

\noindent \textit{\textbf{Objective:}}  
The paper systematically explores data management challenges and potential solutions in agile projects, aiming to provide insights into data management challenges and solutions for both researchers and practitioners.

\noindent \textit{\textbf{Method:}} 
We employed a mixed-methods approach, including a systematic literature review (SLR) to understand the state-of-research followed by a survey with practitioners to reflect on the state-of-practice. The SLR reviewed 45 studies, identifying and categorizing data management aspects along with their associated challenges and solutions. The practitioner survey captured practical experiences and solutions from 32 industry practitioners who were significantly involved in data management to complement the findings from the SLR. 

\noindent \textit{\textbf{Results:}} 
Our findings identified major data management challenges in practice, such as managing data integration processes, capturing diverse data, automating data collection, and meeting real-time analysis requirements. To address these challenges, solutions such as automation tools, decentralized data management practices, and ontology-based approaches have been identified. These solutions enhance data integration, improve data quality, and enable real-time decision-making by providing flexible frameworks tailored to agile project needs.


\noindent \textit{\textbf{Conclusion:}} 
The study pinpointed significant challenges and actionable solutions in data management for agile software development. Our findings provide practical implications for practitioners and researchers, emphasizing the development of effective data management practices and tools to address those challenges and improve project success.
\endgroup
\end{abstract}
\section{Introduction}
\label{sec:Combined Intoduction}
Agile software development approaches prioritize flexibility, rapid iteration, and constant adaptation to change. Its dynamic nature can lead to issues related to how data (product, process, project and operational data) is managed in a project, such as integrating diverse data sources and maintaining data quality despite frequent updates and changes \cite{franch2019quality}. When managing data, agile projects often face problems with data collection and inconsistencies. For example, Fabijan et al. \cite{fabijan2016lack} observed that data collected by different agile teams often remained isolated in large software-intensive organizations, leading to inefficiencies and repeated efforts, as important feedback was not systematically shared across the organization. A product owner mentioned ``\textit{it is all in my head more or less}'', highlighting the issue of unshared, valuable data \cite{fabijan2016lack}.
Furthermore, the dynamic nature of privacy requirements and the frequent updates to user stories exacerbate the challenge, as teams often struggle to keep documentation current and comprehensive. Managing data in agile projects poses significant challenges that must be effectively handled to optimize organizational performance \cite{graetsch2023dealing}. \textbf{This research investigates} data management challenges and solutions in agile software development. We employed a mixed-methods approach using a Systematic Literature Review (SLR) and a survey with practitioners to explore data management challenges and solutions in agile software development.

Several studies have explored data management in agile software development, such as those by Vestues et al. \cite{vestues2022agile} and Graetsch et al. \cite{graetsch2023dealing}. However, these studies focused on specific case studies from individual projects and organizations. For example, Vestues et al. \cite{vestues2022agile} discussed a case study on implementing a data mesh to transition from \textit{centralized} to \textit{distributed} data management, highlighting its associated challenges and benefits. Graetsch et al. \cite{graetsch2023dealing} emphasized the value of real-time data analytics, basing their study on interviews with practitioners to examine the challenges of delivering data-intensive software solutions.
In contrast, \textbf{our SLR examines diverse software development projects across various industries}, addressing multiple data management aspects. We also present implications for agile teams and project delivery. Conducting the SLR allows us to gather and aggregate evidence related to different data management challenges and solutions presented in previous research. Based on the findings from the SLR, which identified the challenges and solutions noted in previous research, we then designed a practitioner survey to better understand the challenges practitioners encounter, pinpoint workable solutions used in industries and uncover additional insights that may not be covered in the existing literature. We used the key challenges and solutions we identified in the SLR to inform our questionnaire structure and questions, which we then distributed to participants.
The survey also offers insightful information about project-specific challenges and solutions practitioners implement to manage data efficiently in agile settings. Combining the results from the SLR with the practitioner survey can ensure an integrated knowledge of data management challenges and solutions from both academic and practical viewpoints. Moreover, the results of this study informed future research directions and managerial implications of data management in agile software development.
Our study aims to answer the following two research questions (RQs):
\begin{itemize}
\item[] \textbf{RQ1}. What are data management challenges in agile software development?\\
\item[] \textbf{RQ2}. What are solutions proposed to address these challenges?
\end{itemize}

The survey conducted among practitioners confirms the majority of \textbf{the findings} from the SLR and leads to new insights. Both the SLR and the survey show that agile teams struggle with managing data, such as integrating and collecting data from diverse sources and maintaining data quality, which hampers collaboration and well-informed decisions in projects. Solutions such as using ontologies, decentralized data management, automation tools, and communication-centric approaches were proposed. The survey emphasizes the effectiveness of automation tools and decentralized data management practices in addressing data integration and quality challenges. Moreover, the survey highlights the specific impact of these challenges on agile teams and the importance of developing clear data management policies and providing team training. Our study reveals that while managerial efforts should prioritize data management policy development, training, and tool adoption, future research should concentrate on improved integration approaches, automated quality assurance, data collection, and real-time analytics.

The remaining sections of this paper are structured as follows: 
Section \ref{sec:background} provides the background on data management and agile methods. We present the SLR in Section \ref{sec:slr} followed by the practitioner survey in Section \ref{sec:survey}. We discuss the study results and present a list of implications and recommendations in Section \ref{sec:discussion} followed by a discussion of threats to validity in Section \ref{Full-ThreatstoValidity}. Finally, the conclusion of this study is presented in Section \ref{sec:conclusion}.

\section{Background}
\label{sec:background}
\subsection{Agile Software Development and Importance of Data Management}
Agile software development aims to make incremental progress by continuously incorporating customer feedback and adapting to change through a flexible and collaborative approach. Agile development encourages a development approach focused on the customer's needs and appreciates the contributions of skilled and empowered teams. Agile methods prioritize people, working software, customer collaboration, and adaptability over rigid processes and planning, as stated in the agile manifesto \cite{beck2001manifesto}. Agile practices have increased the effectiveness and responsiveness of software development to market demands \cite{amajuoyi2024agile}.

Data management plays a crucial role in agile projects. Data management refers to the process of securely, efficiently, and cost-effectively gathering, storing, and utilizing data. It guarantees that data is easily accessible, reliable, and promptly available to consumers. Data management is crucial because it facilitates decision-making, enhances business performance, and enhances customer comprehension and service \cite{data-management-IBM}.

Organizational success is significantly impacted by managing data, which is a crucial resource that helps with decision-making, provides insightful information about customer behavior, and improves operational efficiency \cite{matthies2019towards}. Many forms of data (e.g., business or product data) play important roles in agile development, each impacting the development process and the project's overall success.

\subsection{Types of Data in Agile Software Development} 
Different data types can have different implications on agile processes. We discuss those data types and their possible implications below:

\textit{Business data:} This data type contains details relevant to business decisions, such as forecasting sales, anticipating the need for raw materials, and analyzing consumer behavior \cite{matthies2019towards}. This data type enhances productivity and operational efficiency by facilitating direct feedback from users and enabling quick communication between developers and customers in an agile process. This data facilitates iterative cycles of decision-making in agile processes. Teams can use it to prioritize features according to market trends and customer requirements, ensuring that development efforts align with company objectives \cite{fabijan2016lack}. \textcolor{black}{For example, during the COVID-19 pandemic, population flow data (i.e., aggregated data from mobile phone location tracking or similar systems that show how people move between areas) was used to identify high-risk regions, allowing targeted health measures and better resource planning \cite{huang2021leveraging}.}

\textit{Product Data:} This data type contains details about the software product itself, including design documents and source code. It measures the software product's size, complexity, and design structure,  which all significantly impact software quality \cite{rahman2013and,li1999software,colakoglu2021software}. This data provides metrics that can be used to measure code quality, complexity, and technical debt to inform continuous integration and agile delivery processes. Such metrics are critical to sustaining high levels of software quality and enabling quick releases \cite{rosenkranz2017supporting}. \textcolor{black}{For example, privacy requirements can be integrated into user stories to ensure compliance with the data protection law \cite{canedo2022guidelines}, which helped agile teams assess and implement data anonymization measures and other privacy safeguards during the development process.}

\textit{Process Data:} Process data is about the software development process and includes several elements that could affect how much work is needed to build a software system. Understanding and enhancing the software development lifecycle need the use of process data \cite{rahman2013and,li1999software}. Process data is essential for continual improvement, which agile techniques emphasize. Teams can modify their procedures for improved results by using the insights it offers about process bottlenecks, team performance, and collaboration efficiency \cite{matthies2019towards}. \textcolor{black}{For example, Fabijan et al. \cite{fabijan2016lack} analyzed task handover times between development and testing phases to identify delays and inefficiencies. This data revealed bottlenecks caused by a lack of data sharing, leading to repetitive work and suboptimal decision-making. Addressing these issues streamlined the development process and improved overall team collaboration.}

\textit{Project Data:} This data type focuses on a software project's general health and state, including resources, risks, budgets, and schedules. Project managers frequently use this data to track the status of their work and inform choices about planning and controlling their projects \cite{colakoglu2021software}. Agile project management depends on having real-time project data. It assists in monitoring development, controlling risks, and making sure the project stays within budget and on schedule. This data demonstrates the agile principles of transparency and ongoing feedback \cite{matthies2019towards}. \textcolor{black}{For example, metrics like the mean time between failures (MTBF) have been used to predict and address potential software errors during agile sprints. Such project data supports resource allocation and task prioritization, improving project outcomes \cite{batarseh2018predicting}.}

\textit{Operational data} is data about a company's daily activities. It entails privacy engineering, managing personal data, and deploying and running software in cloud environments \cite{grunewald2021cloud}. Operational data comprises sensitive corporate information, classified documents, and data kept in databases necessary for enterprise operations. Such data is crucial to the day-to-day functioning of the business \cite{min2010practices}. Within the agile framework, operational data facilitates the daily management of projects.
\textcolor{black}{For example, agile methods were applied to create patient registries using electronic health record data for chronic disease management \cite{kannan2017rapid}. This operational data, managed by the organization deploying the system, supported real-time patient monitoring and reporting \cite{kannan2017rapid}. In contrast to operational data, which focuses on supporting daily activities, business data is used for strategic decision-making (e.g., aggregating population flow data during COVID-19 to identify high-risk areas and optimize resource allocation).}

These different data types show how they may be used to better decision-making procedures, encourage client participation, and enhance software development procedures—all of which contribute to an organization's success. 
There are various reasons why data type classification is important:
Agile teams can make more informed decisions when they better understand the various forms of data and their functions. For example, combining process and business data can yield detailed information about customer satisfaction and project performance \cite{fabijan2016lack}.

Categorizing data makes it easier to pinpoint the unique challenges of each type of data. For instance, monitoring operational data is crucial for compliance and security, while guaranteeing the correctness and consistency of product data is crucial for preserving software quality \cite{matthies2019towards}.

Developing targeted solutions for specific challenges is made more accessible by classifying data. When the type of data and its challenges are well-defined, for instance, applying ontologies for data integration or using automated methods for data quality assurance can be done more successfully \cite{barcellos2020towards}.

\subsection{Related Work}

To the best of our knowledge, our study is the first review that specifically investigates data management challenges and solutions in agile software projects. An SLR systematically gathers and examines evidence from multiple studies, creating a comprehensive understanding of data management challenges and solutions in agile software development. Previous studies have mostly focused on separating a set of challenges or solutions, which can be systems-specific. 
For example, Graetsch et al. \cite{graetsch2023dealing} discussed dealing with data challenges when delivering data-intensive software solutions but did not discuss 
specific solutions that have been implemented. Similarly, the solution proposed by Rosenkranz et al. \cite{rosenkranz2017supporting} addresses specific data integration (e.g., harmonizing data from different sources) and data quality challenges (e.g., ensuring accuracy and consistency), highlighting the interconnected nature of these two aspects in effective data management. Another important solution not covered in the previous studies is the use of methods for collecting diverse data through user-centered design  \cite{pater2018advancing}. 

Those diverse challenges and solutions discussed in previous studies highlighted the need for a more comprehensive study to understand the context of those challenges and the relationship between the noted challenges and proposed solutions.  
The review can help provide a better understanding of existing work in data management, identify gaps that need further investigation, and avoid duplication of research efforts. It also offers practitioners a consolidated source of data management challenges and solutions. In the following two sections, we report our SLR study followed by the practitioner survey.

\section{Systematic Literature Review on Data Management}
\label{sec:slr}
As noted in Section \ref{sec:Combined Intoduction}, the first step of our investigation is to conduct an SLR of data management challenges and solutions in agile software development. The review aims to answer the two RQs noted in Section~\ref{sec:Combined Intoduction}, and scope the practitioner survey in Section~\ref{sec:survey}. 

\subsection{Systematic Literature Review Method}
We designed this review by following the SLR guidelines of Kitchenham and Charters \cite{kitchenham2007guidelines} and drew upon works related to agile literature reviews, such as Dikert et al. \cite{dikert2016challenges} and Campanelli and Parreiras \cite{campanelli2015agile}. The review process is presented in Figure \ref{fig:fig-ReviewProcess}. 

\subsubsection{Search Process}
\label{Subsec-SearchProcess}
We used Scopus to search for relevant studies due to its comprehensive indexing of major publications in software engineering, as highlighted in studies by Mour{\~a}o et al. \cite{mourao2020performance} and Carrera-Rivera et al. \cite{carrera2022conduct}. The extensive coverage of Scopus provides access to high-quality, peer-reviewed literature, which is crucial for an in-depth exploration of agile software development. Moreover, data management is a topic at the intersection of multiple disciplines (including software engineering, project management, etc.). Scopus provides more extensive coverage of literature on our study topic as it covers popular publishing venues, including the four major software engineering publication libraries: IEEE Xplore,  ACM Digital Library, SpringerLink, and ScienceDirect. The search covered literature starts from December 2003 (the date of the first study returned by our search string, as there was no start date limit) until October 2023, when we began consolidating the findings to guide the survey.
 
We constructed a search string covering the main keywords of the studies we aim to review. We first identified a set of initial keywords for our search string: \textit{data management, agile, challenges and solutions}. This set of keywords was iteratively refined to review article coverage. We experimented with different search strings until we reached a string that returned relevant results. We had a quasi-gold standard \cite{zhang2011identifying} as a set of studies we knew were relevant and the search string found them all (included in our replication package \cite{fawzy_2024_10597818}).

Before finalizing our search keywords and search string and establishing our inclusion and exclusion criteria, we initiated a pilot search with a basic string (``agile AND data'') in the Scopus database. This initial search generated 9,900 studies, yet a review of a sample of titles and abstracts revealed their lack of relevance to our focus on data management challenges in agile software development, being too broad for our purposes.

To refine our search and secure more pertinent results, we expanded the string by incorporating ``challenge'' and adjusting synonyms like ``obstacle'', ``issue'', and ``problem'' with the OR operator. This refinement reduced the number of results to 3,068 studies. However, a further examination of a new sample of titles and abstracts showed that they were not explicitly related to software development, as our research specifically targeted agile software development, not agile methodologies in a broader sense.

We continued to refine our approach, modifying the search string to ``(agile AND data AND software) AND (challenge OR practice)'' which led to 1,396 studies. Yet, these results still did not focus explicitly on data management. To address this, we further refined the search to ``(agile AND data AND software) AND (challenge OR practice) AND (``data management'')'', achieving more targeted results. However, this specific approach risked missing relevant studies where ``data management'' appeared in the context of its various aspects rather than as an explicit term.

To overcome this, we expanded the search string to include a range of terms associated with data management, resulting in a more balanced set of 181 studies as of the final execution in October 2023. To avoid missing related studies by focusing narrowly on specific terms, we used Boolean operators (OR) among them or included the general term ``data management''. A random review of sample titles and abstracts from these 181 studies demonstrated greater relevance to our research question.

\begin{figure}[H]
    \centering
    \includegraphics[width=0.70\linewidth]{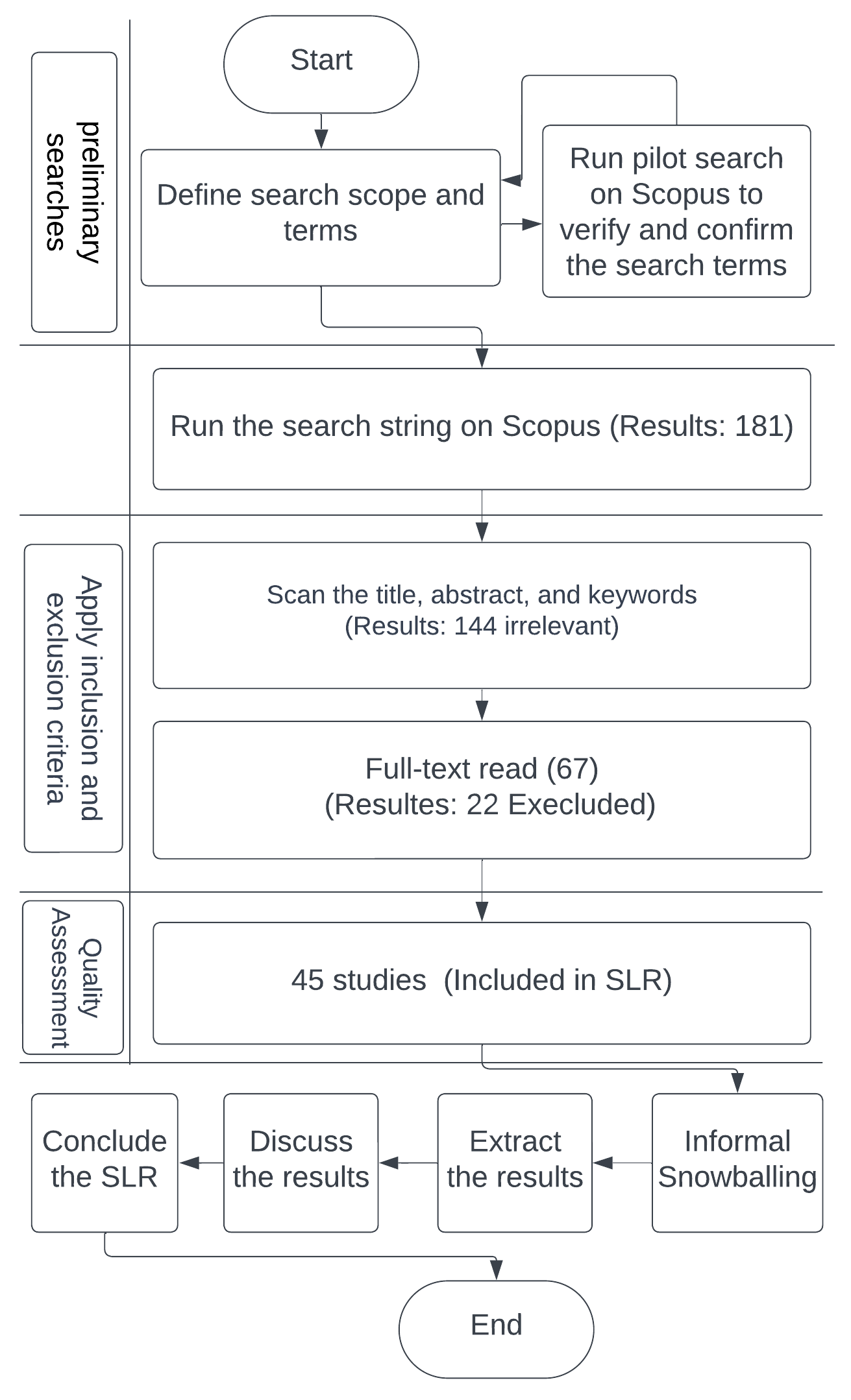}
    \caption{The Systematic Literature Review Process}
    \label{fig:fig-ReviewProcess}
\end{figure}

These 181 studies proceeded to the next stage of review, as illustrated in Figure \ref{fig:fig-ReviewProcess}. Additionally, we conducted a full pilot review process on a sample of these studies. We applied the inclusion and exclusion criteria, performed quality assessments (see Section \ref{Sub-QA}), and extracted the data to ensure the final selection aligned with our research objectives.

In the search string, we aim to cover all two aspects of the study, i.e., \textit{agile software development} and \textit{data management}, to capture studies that focus on data management (aspects, issues, challenges, and solutions) in agile software development.  
Our final search string (which covers the aspects above) is shown below:

\begin{mytextbox}
\texttt{(agile AND software AND data) AND (challenge OR practice) AND ("data quality" OR "data management" OR "data security" OR "data governance" OR "data integration" OR "data storage" OR "data privacy" OR "data access" OR "data analytics" OR "data validation" OR "data capture" OR "data modeling" OR "data virtualization" OR "data cataloging" OR "data Versioning" OR "data monitoring" OR "data transformation" OR "data archiving" OR "backup" OR "disaster recovery" OR "data-driven")}
\end{mytextbox}

To ensure the completeness and reliability of our study selection, we adopted an informal snowballing approach as a supplementary validation step. Specifically, we examined the references of a subset of the top nine cited studies, treating them as a quasi-gold standard \cite{zhang2011identifying}, i.e., a set of highly relevant studies. This backward snowballing approach allowed us to cross-check our dataset and confirm that no key studies were missed due to search string limitations. After reviewing these references, we found that no additional studies met our inclusion criteria, reinforcing the robustness and comprehensiveness of our selection process. We conducted this search before extracting the results (see Figure \ref{fig:fig-ReviewProcess}).

\subsubsection{Inclusion and Exclusion Criteria}
\label{Subsec:Selection and Exclusion Criteria}

 We selected studies that discuss at least one aspect of data management in the context of agile software development (including challenges and solutions). We applied the following exclusion criteria:
\begin{enumerate}
\item  \textcolor{black}{Studies that are not available in full text (e.g., abstracts only).}
\item \textcolor{black}{Studies that are not published in English.} 
\item Studies that do not specifically mention agile software development or agile methodologies.
    \item Duplicate studies to ensure the uniqueness of each data point.
     \item Studies that focus exclusively on technical or other aspects of software development without demonstrating a clear link to data management challenges in agile software development.
      \item Non-research materials such as tutorials, books, or workshop summaries.
    \item Opinion pieces, editorials, or non-peer-reviewed articles to ensure the academic rigor of the review.
  \end{enumerate}

\subsubsection{Filtering of Studies}
We screened all 181 studies and examined all titles, abstracts, and keywords used. \textcolor{black}{The first author conducted the initial screening, applying the inclusion and exclusion criteria to ensure relevance. The second co-author cross-validated the screening to verify consistency and accuracy.} By applying the inclusion and exclusion criteria, we filtered out 114 irrelevant studies, leaving 67 relevant studies. To ensure that these studies precisely matched our research questions and scope, we conducted a full-text review, applying the same criteria. \textcolor{black}{The full-text review followed the same structured approach: the first co-author conducted the initial review, and the second co-author cross-validated the results.} This review led to the exclusion of 22 studies from the initial 67. Reasons for exclusion included factors such as language barriers (e.g., one study was written in a non-English language, with only the abstract in English), content type (e.g., one was a tutorial, another an event), or lack of depth or relevance to our scope. 

\textcolor{black}{The filtration process was conducted independently by the first co-author in the initial phase, followed by cross-validation by the second co-author. Any disagreements were resolved through discussions and consensus among all four authors, ensuring transparency and rigor. 
} After applying the filtration process, we included a total of 45 studies (see Section \ref{Sub-QA}).

\subsubsection{Data Extraction}
\label{sub:datExtraction}

We utilized a combination of thematic analysis and content analysis to synthesize data from the studies. Thematic analysis \cite{braun2006using} was used to categorize, analyze, and report themes (data management aspect, challenge, and solution) within the qualitative data extracted from the 45 studies. Initially, a subset of studies (9 studies) were reviewed to identify general data management aspects. This provides a structured understanding of the various aspects of data management within agile software development. Additionally, this foundational knowledge helps identify specific challenges and solutions associated with each aspect, thereby addressing RQs.
We categorized the extracted data from the studies (all those items were extracted as free text) as follows: 
\begin{itemize}
\item study objective, 
\item data management aspect, 
\item challenges identified (RQ1), and 
\item suggested solutions (RQ2)
\end{itemize}

\textcolor{black}{Our data extraction process was carried out collaboratively to ensure accuracy and reliability. The first co-author performed the initial data extraction, while the second co-author cross-validated the extracted data to ensure consistency and alignment with the research objectives. The last two co-authors reviewed the extracted data, offering insights and refinements to enhance the comprehensiveness of the results.}

\textcolor{black}{Subsequently, any disagreements in the extracted data were resolved through discussions among all authors until a consensus was reached.} Relevant data was retrieved and organized by careful iterative reading through the complete texts of 45 included studies for each research question. Data management challenges (RQ1) were identified based on descriptions of problems faced in managing data within agile projects, as documented in the reviewed studies. The solutions (RQ2) were extracted by noting specific strategies, tools, or methods proposed or implemented to address the identified challenges. For example, thematic analysis revealed several challenges associated with data integration (see Section \ref{sub-DICS}), such as data harmonization, semantic heterogeneity, data transformation, and extraction. The identified solutions were also thematically categorized and presented (see Table \ref{tab:tab-DataIntegrationChallengesandSolutions}). An iterative process was used to review and validate the extracted data. 
To ensure accuracy and consistency, the two coders discussed the extracted data from a sample of studies and compared the outcomes. They also discussed any differences to settle any disagreements in classification. We also employed content analysis \cite{krippendorff2018content} to analyze studies based on the data management aspects they addressed. Additionally, we quantified the number of studies discussing each aspect to identify the challenges and solutions that were covered the most and least. For example, the studies were classified into 15 data management aspects (see Section \ref{sec:data-management-aspects}), listed these aspects, and showed the distribution of studies across these aspects. In addition, the data types discussed in the studies are provided in Table \ref{tab:DataTypes}. We provide a dataset of all the included studies we have extracted and our detailed analysis in a replication package \cite{fawzy_2024_10597818}.

\textcolor{black}{Both thematic and content analysis were important to ensure a comprehensive and balanced synthesis of the findings. Thematic analysis allowed us to gain deep insights into the studies by identifying recurring patterns and themes. For instance, it highlighted that challenges with data harmonization often overlap with issues in ensuring data accuracy and consistency due to the lack of standardization in data integration pipelines. On the other hand, content analysis enabled us to quantify the prevalence of these challenges across the studies, revealing their relative significance, such as data integration being addressed in 19 studies  (see Section \ref{DMPRZ}). This mixed-methods approach combined the depth of qualitative insights with the validation of quantitative evidence. Integrating qualitative and quantitative methods enables a deeper understanding of complex phenomena by combining the detailed exploration of qualitative data (e.g., uncovering relationships between challenges) with the ability of quantitative data to validate findings across a broader range of studies (e.g., identifying widely addressed challenges) \cite{fetters2013achieving}. The prioritization of data management challenges (see Section \ref{DMPRZ}) is an example of the complementary contributions of both methods.}
\begingroup 
\color{black} 
The following are the combined thematic and content analysis steps that we follow:
\begin{itemize}
\item \textit{Understanding the Data:} We first familiarized ourselves with the data by reading over the studies included in the SLR after applying the filtration process. 
.
At this stage, content analysis focused on identifying broad categories or topics within the filtered dataset 
noted for their frequency and relevance to inform the next steps in analysis.

\item \textit{Generating Initial Codes:} We have iteratively read each of the included studies to identify the possible codes and themes. The first author developed the initial codes, which the second author then reviewed for consistency. 
These initial codes highlighted key patterns in the data and formed the basis for further analysis. We employed  content analysis to quantify mentions of key terms or challenges. 
\item \textit{Searching for Themes:} Related codes were grouped into broader themes. For instance, ``semantic heterogeneity'' and ``data harmonization'' were grouped under the theme ``Data Integration''. Thematic grouping was supported by calculating the prevalence of codes. For example, the theme ``Data Integration'' emerged as the most frequently discussed aspect, appearing in 19 studies. This quantification ensured that identified themes were representative of the dataset.
\item \textit{Reviewing Themes:} The relationships between themes were examined. For example, challenges with data integration often overlapped with data quality issues, reflecting their interconnected nature in agile projects. The data was cross-checked against the initial studies to ensure accuracy and consistency. The extracted quantitative data from content analysis validated thematic findings by showing overlap in study coverage. For example, several studies discussed both ''data integration`` and ''data quality``, underscoring the interconnected nature of these challenges. This helped ensure the themes were comprehensive and grounded in evidence.
\item \textit{Defining and Refining Themes:} Each theme was clearly defined and named. For example, ``Data Integration'' was described as the process of combining data from various sources to form a unified dataset.
Content analysis helped quantify the prevalence of the challenges across the studies. For instance, the ``Data Integration'' theme was supported by 4 studies discussing ``data harmonization and interoperability'' and 4 studies addressing ``semantic heterogeneity'', as shown in Table \ref{DMPRZ}. Similarly, the ``Data Quality'' theme was supported by 5 studies focusing on ``ensuring data accuracy and consistency''. 

\end{itemize}

\endgroup

\subsubsection{Quality Assessment}
\label{Sub-QA}
Throughout our review, we have conducted comprehensive quality checks to ensure rigor, integrity, and relevance. This process began with examining selected samples of search string results, aligning them carefully with our established inclusion and exclusion criteria to ensure the focus and relevance of our study. Furthermore, we conducted an in-depth quality assessment of the 45 studies, employing a streamlined, three-point scoring system across six key criteria \cite{QualityAssessment}: \textit{Clarity of Objectives},\textit{ Appropriateness of Methodology}, \textit{Adequacy of Data Analysis}, \textit{Relevance to Research Questions}, \textit{Rigor of Data Collection}, and \textit{Quality of Reporting}. Each criterion is rated 1 (adequate), 2 (good), or 3 (excellent), depending on the study's clarity, methodological soundness, analytical rigor, relevance, data collection thoroughness, and quality of reporting. We calculate the total score for each study by summing the scores from all criteria, with the maximum possible score being 18. The studies are then categorized according to their total scores, with 7-11 indicating adequate quality, 12-15 indicating good quality, and 16-18 representing excellent quality. This meticulous approach ensures a comprehensive and nuanced assessment of each study's quality, aligning with our goal to exclude low-quality research and thereby maintaining the integrity and reliability of our findings. The quality assessment confirms that the 45 studies are of good quality (the scores of these 45 studies are all above 12). Details of the quality assessment, including the scoring system, can be found in the replication package \cite{fawzy_2024_10597818}.

\subsection{Results}
\label{sec:results}
This section provides a detailed analysis of various data management aspects and answers the two RQs. First, we define four data management aspects, as identified in our SLR in Section \ref{sec:data-management-aspects}. Structured based on these aspects, we then discuss the challenges (RQ1) and solutions (RQ2) associated with each aspect in detail (see Sections \ref{sub-DICS} -- \ref{sub-DACS}). Note that we discuss the challenges and solutions together, as this will provide readers with a better context of each challenge and the related solution as discussed in the associated study. \textcolor{black}{We provided a definition for each of those challenges and solutions, which emerged from analyzing and synthesizing the studies included in the SLR}.

\subsubsection{Data Management Aspects}
\label{sec:data-management-aspects}
Below, we present the definitions of those four top data management aspects. We have provided definitions of the other 11 data management aspects, along with a summary of their challenges and solutions, in the replication package \cite{fawzy_2024_10597818}. 

\textit{\textbf{Data Integration}} refers to the process of combining data from various sources and systems into a unified and consistent view. This consolidated data is then utilized for analysis, reporting, and decision-making purposes, as referred to in the studies that discussed the data integration aspect in Table \ref{tab:results-classification32}.

\textit{\textbf{Data Collection}}
is an activity that involves gathering information. This information is subsequently utilized for analysis and decision-making, as mentioned in the studies discussed data collection aspect in Table \ref{tab:results-classification32}.

\textit{\textbf{Data Quality}}
is defined by the accuracy, completeness, and consistency of a dataset. The importance of high-quality data is underscored by its critical role in effective data analysis and decision-making processes. This is according to the studies that discussed the data quality aspect in Table \ref{tab:results-classification32}.

\textit{\textbf{Data Analysis}} is a comprehensive activity that inspects, cleans, transforms, and models data to unearth valuable information, obtain conclusions, and strengthen decision-making. This activity necessitates the management of diverse data types, including structured and unstructured data, and requires diverse techniques and tools. This is according to the studies that discussed the data analysis aspect in Table \ref{tab:results-classification32}.

\begin{figure}
    \centering
    \includegraphics[width=1\linewidth]{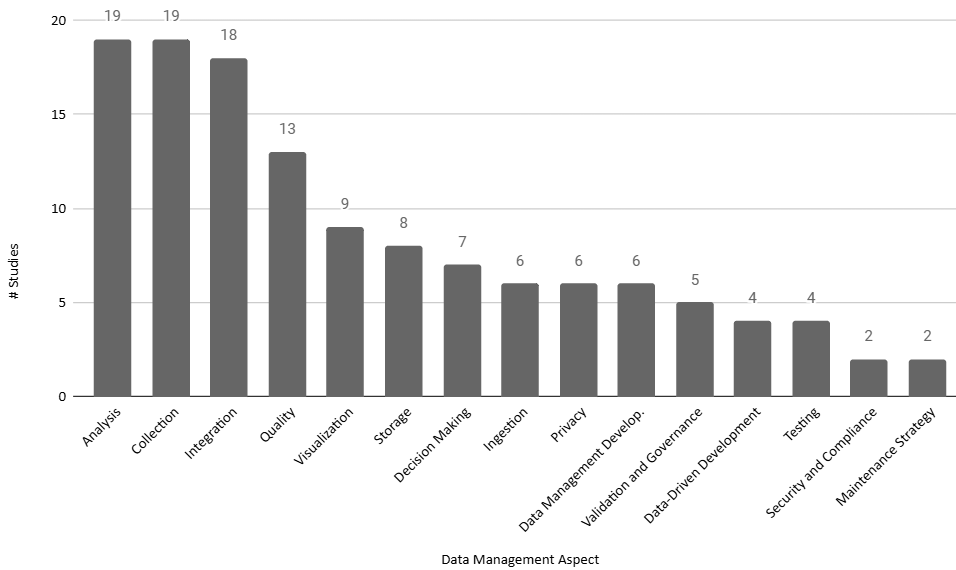}
    \caption{Total Number of Studies Discussing Challenges and Solutions Associated with Fifteen Aspects of Data Management}
    \label{fig:Graph-AspectsandNumberofStudies}
\end{figure}

\textbf{Distribution of Studies across Data Management Aspects:}
We classified each of the 45 studies based on the data management aspects they address. Table \ref{tab:results-classification32} lists all 15 data management aspects across studies. Figure \ref{fig:Graph-AspectsandNumberofStudies} shows the distribution of studies across the fifteen data management aspects we identified.

\begin{sidewaystable}[]
\centering
\caption{Classifications of Studies Across Data Management Aspects}
\resizebox{\columnwidth}{!}{%
\begin{tabular}{lllllllllllllllllllll}
\toprule

 & \textbf{Data Management Aspect}   &      &                                 &       \textbf{ Ref.  }                      &                                   &                             &                                 &                                   &                               &                             &                             &                                   &                                 &                             &                                   &                           &                           &                                 &                                 \\
 \midrule
1  & Data Integration               & \cite{chen2016agile}        & \cite{harper2014agile}         & \cite{dos2021ontology}          & \cite{chhillar2019act}         & \cite{upender2005staying}         & \cite{abdallah2022towards}  & \cite{barcellos2020towards}     & \cite{rix2016agile}               & \cite{vestues2022agile}       & \cite{spengler2020enabling} & \cite{kannan2017rapid}      & \cite{hofer2020new}               & \cite{rosenkranz2017supporting} & \cite{little2004adaptive}   & \cite{schuttler2021journey}       & \cite{dharmapal2016big}   & \cite{dursun2014workflow} & \cite{vogt2023implementing}     &                                 \\
2  & Data Storage                   & \cite{chen2016agile}        & \cite{harper2014agile}         & \cite{barbala2023data}          & \cite{barcellos2020towards}    & \cite{schuttler2021journey}       & \cite{dharmapal2016big}     & \cite{huang2021leveraging}      & \cite{dursun2014workflow}         &                               &                             &                             &                                   &                                 &                             &                                   &                           &                           &                                 &                                 \\
3  & Data Validation and Governance & \cite{chen2016agile}        & \cite{kaur2020dialogue}                 & \cite{martinez2019continuously} & \cite{hofer2020new}            & \cite{harriman2004emergent}       &                             &                                 &                                   &                               &                             &                             &                                   &                                 &                             &                                   &                           &                           &                                 &                                 \\
4  & Data Quality                   & \cite{franch2019quality}    & \cite{pater2018advancing}      & \cite{svensson2019unfulfilled}  & \cite{harper2014agile}         & \cite{batarseh2018predicting}     & \cite{ambler2008gets}       & \cite{martinez2019continuously} & \cite{lehtonen2017visualizations} & \cite{spengler2020enabling}   & \cite{kannan2017rapid}      & \cite{hofer2020new}         & \cite{rosenkranz2017supporting}   & \cite{little2004adaptive}       &                             &                                   &                           &                           &                                 &                                 \\
5  & Data Ingestion                 & \cite{chen2016agile}        & \cite{dos2021ontology}         & \cite{dautov2022towards}        & \cite{das2015towards}          & \cite{lehtonen2017visualizations} & \cite{huang2021leveraging}  &                                 &                                   &                               &                             &                             &                                   &                                 &                             &                                   &                           &                           &                                 &                                 \\
6  & Data Collection                & \cite{pater2018advancing}   & \cite{matthies2019towards}     & \cite{svensson2019unfulfilled}  & \cite{chhillar2019act}         & \cite{dos2021ontology}            & \cite{fabijan2016lack}      & \cite{upender2005staying}       & \cite{das2015towards}             & \cite{batarseh2018predicting} & \cite{barbala2023data}      & \cite{kaur2020dialogue}              & \cite{matthies2020playing}        & \cite{martinez2019continuously} & \cite{olsson2018challenges} & \cite{lehtonen2017visualizations} & \cite{dharmapal2016big}   & \cite{lin2018towards}     & \cite{huang2021leveraging}      & \cite{fagarasan2023integrating} \\
7  & Data Security and Compliance  & \cite{upender2005staying}   & \cite{min2010practices}        &                                 &                                &                                   &                             &                                 &                                   &                               &                             &                             &                                   &                                 &                             &                                   &                           &                           &                                 &                                 \\
8  & Data Privacy                   & \cite{canedo2022guidelines} & \cite{grunewald2021cloud}      & \cite{vestues2022agile}         & \cite{barbala2023data}         & \cite{olsson2018challenges}       & \cite{schuttler2021journey} &                                 &                                   &                               &                             &                             &                                   &                                 &                             &                                   &                           &                           &                                 &                                 \\
9  & Data Analysis                  & \cite{chen2016agile}        & \cite{harper2014agile}         & \cite{dos2021ontology}          & \cite{matthies2019towards}     & \cite{svensson2019unfulfilled}    & \cite{chhillar2019act}      & \cite{das2015towards}           & \cite{batarseh2018predicting}     & \cite{matthies2020playing}    & \cite{olsson2018challenges} & \cite{barcellos2020towards} & \cite{lehtonen2017visualizations} & \cite{rix2016agile}             & \cite{dharmapal2016big}     & \cite{lin2018towards}             & \cite{dursun2014workflow} & \cite{hamer2023students}  & \cite{fagarasan2023integrating} & \cite{fabijan2016lack}                                 \\
10 & Data Visualization             & \cite{harper2014agile}      & \cite{svensson2019unfulfilled} & \cite{martinez2019continuously} & \cite{rix2016agile}            & \cite{kannan2017rapid}            & \cite{lin2018towards}       & \cite{hamer2023students}        & \cite{fagarasan2023integrating}   & \cite{lehtonen2017visualizations}            &                             &                             &                                   &                                 &                             &                                   &                           &                           &                                 &                                 \\
11 & Data-Driven Decision Making    & \cite{pater2018advancing}   & \cite{dos2021ontology}         & \cite{matthies2019towards}      & \cite{svensson2019unfulfilled} & \cite{barcellos2020towards}       & \cite{bosch2019towards}     & \cite{lin2018towards}           &                                   &                               &                             &                             &                                   &                                 &                             &                                   &                           &                           &                                 &                                 \\
12 & Data-Driven Development     & \cite{barbala2023data}      & \cite{alsaadi2022data}         & \cite{olsson2018challenges}     & \cite{dursun2014workflow}      &                                   &                             &                                 &                                   &                               &                             &                             &                                   &                                 &                             &                                   &                           &                           &                                 &                                 \\
13 & Data Testing:                  & \cite{upender2005staying}   & \cite{batarseh2018predicting}  & \cite{ambler2008gets}           & \cite{little2004adaptive}      &                                   &                             &                                 &                                   &                               &                             &                             &                                   &                                 &                             &                                   &                           &                           &                                 &                                 \\
14 & Data maintenance strategy      & \cite{ambler2008gets}       & \cite{vestues2022agile}        &                                 &                                &                                   &                             &                                 &                                   &                               &                             &                             &                                   &                                 &                             &                                   &                           &                           &                                 &                                 \\
15 & Data Management Development    & \cite{harriman2004emergent} & \cite{jenness2018lsst}         & \cite{chung2006bridging}        & \cite{dursun2014workflow}      & \cite{chen2016agile}              & \cite{chhillar2019act}      &                                 &                                   &                               &                             &                             &                                   &                                 &                             &                                   &                           &                           &                                 &      \\          
\bottomrule
\end{tabular}}
\label{tab:results-classification32}
\end{sidewaystable}

\subsubsection{Overlapping Data Management Aspects}
\label{Sec:overlab}

\begin{figure}
    \centering
    \includegraphics[width=.90\linewidth]{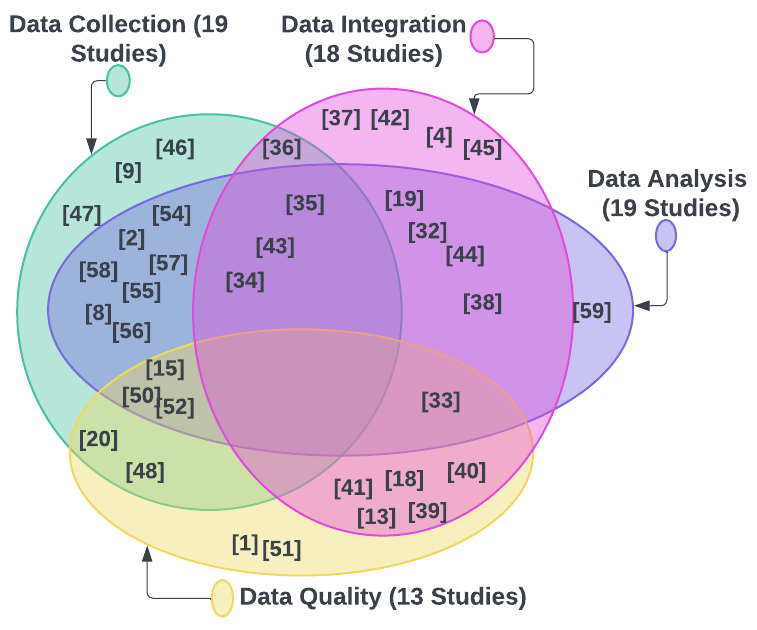}
\caption{Overlapping Studies Between the Key Focus Aspects in Data Management}
    \label{fig:Graph-Venn}
\end{figure}

Figure \ref{fig:Graph-Venn} illustrates the connections between various data management aspects in agile software development, revealing that addressing a single challenge can resolve other connected challenges. This insight underscores the importance of implementing integrated and comprehensive solutions to enhance the effectiveness and efficiency of agile projects.

\textbf{Data Collection and Analysis:}
We observed that the most significant overlap is between data collection and data analysis (as shown in Figure \ref{fig:Graph-Venn}). For example, the Analytics-Driven Testing (ADT) process proposed by Batarseh and Gonzalez \cite{batarseh2018predicting} addresses the challenges in both data collection (unstructured data) and analysis (extracting insights from unstructured data), showcasing the link between these aspects in agile software development and decision-making. Another example is the solution provided by Matthies and Hesse \cite{matthies2019towards}, which addresses both the data collection challenges (variety and volume of sources) and the data analysis challenges (integration and interpretation of diverse data). It demonstrates how practical tool usage can bridge the gap between collecting and analyzing a wide range of data to derive actionable software development insights.

\textbf{Data Integration and Analysis:}
The studies in Figure \ref{fig:Graph-Venn} discussed challenges and solutions related to data integration and analysis, underscoring their role in agile project value delivery, characterized by swift, flexible responses to change. For example, the solution proposed by Chen et al. \cite{chen2016agile} (adopting an architecture-centric approach) addresses the challenges in both data integration (combining diverse data types from Enterprise Data Warehouses (EDWs) and Not Only SQL (NoSQL) systems) and data analysis (applying real-time analytics), while embedding these processes within an agile framework. This approach supports agile's flexible data structure needs for continuous delivery. The interplay between data integration and analysis in agile development is essential for enabling prompt, informed decision-making, adapting swiftly to changes, and maintaining flexible development.

\textbf{Data Integration and Data Quality:}
The studies depicted in Figure \ref{fig:Graph-Venn} discussed challenges and solutions related to data integration and data quality, showing that addressing data integration challenges (Section \ref{sub-DICS}), inherently resolves data quality issues (Section \ref{sub-DQCS}). For example, the solution proposed by Spengler et al. \cite{spengler2020enabling} (cloud-based platform with a generic agile data-loading pipeline) integrates data from distributed and heterogeneous sources. Additionally, this pipeline is designed to automatically detect the syntax and format of input data, handle different encodings, and manage missing and duplicate data. Another example, the solution proposed by Rosenkranz et al. \cite{rosenkranz2017supporting} addresses the challenges in both data integration (harmonizing data from different sources) and data quality (ensuring accuracy and consistency), highlighting the interconnected nature of these two aspects in effective data management.

\subsubsection{\textcolor{black}{Data Management Challenges Prioritization} \label{DMPRZ}}
\textcolor{black}{
We prioritized the data management challenges based on their \emph{prevalence} in reviewed studies and their \emph{potential impact} on agile workflows (see Table \ref{tab:PrioritizedDMC}).}

\textcolor{black}{Prevalence was determined by counting the number of studies discussing each challenge. 
This approach aligns with systematic review guidelines \cite{kitchenham2007guidelines}, which prioritize frequently discussed issues as indicators of their importance. For example, as shown in Table \ref{tab:PrioritizedDMC}, data harmonization and interoperability challenges appeared in 4 studies, highlighting its recurring relevance across domains. Its frequent mention underscores its important role in enabling consistency and seamless integration of diverse systems, a common need across agile projects. Challenges like this, which were discussed extensively, demonstrate their widespread recognition and applicability in practical agile environments.} 

\textcolor{black}{The potential impact was assessed based on how each challenge influences agile practices, such as sprint planning, iterative development, and cross-team collaboration. We categorized impact as follows:}
\begin{itemize}
    \item \textcolor{black}{High: Challenges directly disrupting these practices, such as delaying decision-making or impeding team alignment, were rated high impact. }
    \item \textcolor{black}{Medium: In contrast, those with less severe effects were rated as medium impact. For example, ensuring data accuracy and consistency challenge was identified as having a high impact in 5 studies 
, as inaccurate or inconsistent data can mislead sprint planning and decision-making, causing project delays and inefficiencies. In contrast, while important, Semantic Heterogeneity was rated as medium impact because its effects can often be mitigated with mechanisms such as the use of ontologies. 
}

\end{itemize}

\textcolor{black}{Note that all of the identified challenges were found to have at least medium significance due to their impact on agile workflows. No challenges were categorized as "low impact," reflecting the focus on issues with meaningful implications for agile practices. This highlights the importance of data management in agile environments, where even moderate challenges can significantly affect project outcomes.}

\begin{table}[]
\centering
\begingroup 
\color{black} 
\caption{\textcolor{black}{Prioritized Data Management Challenges 
}}
\label{tab:PrioritizedDMC}
\resizebox{\columnwidth}{!}{%
\begin{tabular}{lll}
\toprule
\textbf{Data Management Challenges}          & \textbf{\begin{tabular}[c]{@{}c@{}}\# Studies\end{tabular}} & \textbf{Potential Impact on Agile Workflows}                     \\
\midrule
\textbf{Data Integration}           &                                                                                     &                                                        \\
\midrule
Data Harmonization and Interoperability& 4                                                                                   & High – Essential for cross-team   workflows            \\
Semantic Heterogeneity                         & 4                                                                                   & Medium – Critical for accurate   decision-making       \\
\begin{tabular}[c]{@{}c@{}}Data Transformation and Extraction \end{tabular}& 4                                                                                   & High – Impacts real-time data   usability              \\
Managing Data Integration                      & 3                                                                                   & Medium – Addresses   cross-functional coordination     \\
Diverse and Decentralized Data Sources         & 3                                                                                   & High – Influences accessibility and  consistency \\
\midrule

\textbf{Data Collection}            &                                                                                     &                                                        \\
\midrule
Capturing Diverse Data                         & 3                                                                                   & High – Impacts completeness of   agile decisions       \\
Comprehensive Data Collection                  & 3                                                                                   & Medium – Enables broader data   applicability          \\
Data Sharing and Collaboration                 & 2                                                                                   & Medium – Improves team alignment                       \\
Informative Data Collection                    & 2                                                                                   & Medium – Supports specific   project needs             \\
\midrule

\textbf{Data Quality }               &                                                                                     &                                                        \\
\midrule
Ensuring Data Accuracy and Consistency         & 5                                                                                   & High – Critical for decision   reliability             \\
Missing Quality Data                           & 3                                                                                   & High – Affects agile sprint   outputs                  \\
Inadequate Data Quality Management             & 3                                                                                   & Medium – Increases the risk of   errors                \\
Data Quality Standardization                   & 2                                                                                   & Medium – Facilitates uniform   data processes          \\
\midrule

\textbf{Data Analysis}              &                                                                                     &                                                        \\
\midrule
Analyzing Large and Complex Data               & 5                                                                                   & High – Influences scalability   and insights           \\
Analyzing Semantic Heterogeneity Data          & 3                                                                                   & Medium – Ensures semantic   alignment                  \\
Efficient Data Analysis and Visualization      & 3                                                                                   & High – Impacts decision-making   speed                 \\
Real-Time Data Analytics and Decision Making   & 2                                                                                   & High – Supports adaptive project   actions             \\
Selection of Appropriate Analytical Techniques & 2                                                                                   & Medium – Improves analysis   relevance      \\
\bottomrule
\end{tabular}%
}
\endgroup 
\end{table}

\subsubsection{Data Integration Challenges and Solutions}
\label{sub-DICS}

We have identified several challenges associated with data integration in agile software development. These challenges and their proposed solutions are discussed below. A summary of these challenges and their corresponding solutions are presented in Table \ref{tab:tab-DataIntegrationChallengesandSolutions}.

\begin{table}[]
\caption{\textcolor{black}{Summary of Data Integration Challenges and Solutions identified in the SLR.}}
\centering
\resizebox{\columnwidth}{!}{%
\begin{tabular}{llll}

\toprule
\textbf{Challenge}     & \textbf{Solution}                                       & \textbf{Solution Status}      & \textbf{Ref.}                                    \\
\midrule
\multirow{4}{*}{\begin{tabular}[l]{@{}l@{}} Data Harmonization \\ and Interoperability \end{tabular}} & Development of ontologies & Proposed                      & \cite{abdallah2022towards}      \\
                                        & Cloud-based platform with agile data-loading pipeline                    & Implemented     & \cite{spengler2020enabling}     \\
                                        & Agile workflow for integrating data              & Proposed        & \cite{vogt2023implementing}     \\
                                        & Translating the datasets' metadata into different formats& Implemented     & \cite{schuttler2021journey}     \\
\midrule
\multirow{4}{*}{Semantic Heterogeneity}                  & Communication-centric approach with agile methods                        & Implemented    & \cite{rosenkranz2017supporting} \\
                                        & Development of ontology-based approach& Proposed                & \cite{dos2021ontology}          \\
                                        & Development of domain ontologies& Implemented     & \cite{barcellos2020towards}     \\
                                        & Implementing a modular and agile framework        & Implemented    & \cite{rix2016agile}             \\
\midrule
\multirow{4}{*}{\begin{tabular}[l]{@{}l@{}} Data Transformation \\ and Extraction \end{tabular}}      & The implementation of advanced ETL procedures& Implemented     & \cite{hofer2020new}             \\
                                        & Data visualization and federation layer                                  & Proposed                      & \cite{dursun2014workflow}       \\
                                        & Replicable ETL process                                  & Implemented    & \cite{kannan2017rapid}          \\
                                        & Shift to advanced analytic platforms                                     & Proposed                     & \cite{dharmapal2016big}         \\
\midrule
\multirow{3}{*}{Managing Data Integration}               & Adoption of `data mesh' approach                                         & Proposed           & \cite{vestues2022agile}         \\
                                        & Adoption of the Microsoft Solutions Framework (MSF)                      & Implemented                   & \cite{little2004adaptive}       \\
                                        & Architecture-centric approach with AABA methodology                      & Implemented     & \cite{chen2016agile}            \\
\midrule
\multirow{3}{*}{\begin{tabular}[l]{@{}l@{}} Diverse and Decentralized \\ Data Sources\end{tabular}}   & Developing a common product line architecture & Implemented                   & \cite{harper2014agile}          \\
                                        & Automated Continuous Quality (ACQ) Metrics dashboard&\multirow{1}{*}{\begin{tabular}[l]{@{}l@{}} Partially\\implemented\end{tabular}} & \cite{chhillar2019act}          \\
                                        &\\
                                        &Ref in Sec. \ref{sub-DCCS}       & \multirow{1}{*}{\begin{tabular}[l]{@{}l@{}}Partially\\implemented\end{tabular}}                            & \cite{upender2005staying} \\
                                         &\\
\bottomrule
                                        
\end{tabular}
}
\label{tab:tab-DataIntegrationChallengesandSolutions}

\end{table}

\noindent\textbf{Data Harmonization and Interoperability:}
\textcolor{black}{ Data harmonization ensures consistency across diverse sources, while interoperability enables seamless data exchange between systems. In agile development, these processes are crucial for integrating data across teams and tools amid frequent iterations and dynamic requirements. Without them, teams face inefficiencies like redundant efforts, fragmented datasets, and poor data alignment, hindering decision-making and project agility.}
Abdallah and Fan \cite{abdallah2022towards} discussed the challenge of harmonization and interoperability between heterogeneous Maintenance Repair and Overhaul (MRO) information systems, which is a critical issue in integrating maintenance records over the lifespan of an aircraft. Spengler et al. \cite{spengler2020enabling} discussed the challenge of integrating data from distributed and heterogeneous sources at the technical, structural, and semantic levels. V{\o}gt et al. \cite{vogt2023implementing} addressed incorporating new climate data into local government processes. Sch{\"u}ttler et al. \cite{schuttler2021journey} focused on coordinating the definition and integration of datasets to ensure interoperability and avoid the development of parallel structures, necessitating alignment with other research infrastructures like the European Biobanking and BioMolecular Resources Research Infrastructure - European Research Infrastructure Consortium (BBMRI-ERIC) and the German Medical Informatics Initiative (MII), which is a national initiative aimed at improving medical research and healthcare by facilitating the sharing and integration of data across various medical institutions.

\textbf{To address the data harmonization and interoperability challenges} Abdallah and Fan \cite{abdallah2022towards} proposed a solution, which is the development of ontologies to semantically integrate heterogeneous maintenance records of MRO systems, enabling consistent understanding and data exchange across different systems. This approach, known as Ontology-Based Data Integration (ODBI), is supported by the Agile Development for Ontology-Based Applications (ADOBA) methodology, which combines ontology development with application integration in an agile framework. The solution was conceptualized and detailed within the framework of the ADOBA methodology. However, Abdallah and Fan \cite{abdallah2022towards} indicated that further validation is necessary to assess the accuracy and applicability of the methodology in developing ontology-based applications. Spengler et al. \cite{spengler2020enabling} presented a cloud-based platform as a solution that includes a generic agile data-loading pipeline. This pipeline supports deploying and customizing clinical and translational data warehousing solutions, specifically i2b2 and tranSMART. It addresses the need for technical and medical expertise to set up secure data warehousing systems and manage multiple warehouse instances for various data-driven research projects. The platform simplifies the complex installation process and enables rapid instantiation of new instances of data warehousing systems, supporting both i2b2 and tranSMART. Additionally, this pipeline is designed to automatically detect the syntax and format of input data, handle different encodings, and manage missing and duplicate data, thus significantly reducing the efforts required for data cleansing and preprocessing. The solution was implemented and evaluated.

V{\o}gt et al. \cite{vogt2023implementing} proposed the development of an agile workflow to integrate climate data into the administrative process of urban land use planning, illustrated through the City of Constance as a case study. This process involves incorporating the Advanced Municipal Climate Data Store (AMCDS) toolbox to facilitate data-informed decision-making across various activities, including analyzing the current situation and its effects, devising measures, testing and adapting them, and monitoring and assessing progress. The solution was developed and proposed, and the next steps involved presenting the workflow to administrative staff in participatory and co-creative workshops to gather feedback, which would then be integrated into an adjusted workflow. Sch{"u}ttler et al. \cite{schuttler2021journey} proposed a solution to ensure the interoperability of the collected data by closely coordinating the definition and integration of datasets with MII. This process included translating the metadata of the sample, donor, and biobank datasets into HL7 FHIR profiles (Health Level Seven Fast Healthcare Interoperability Resources), which standardized the relevant data within the established IT infrastructure and made them more accessible to query. Furthermore, the collaboration with BBMRI-ERIC resulted in significant synergies, enabling the quick technical integration of tools within the IT infrastructure, leading to international visibility and a high degree of interoperability. The solution was implemented and evaluated.\\

\noindent \textbf{Semantic Heterogeneity:}
\textcolor{black}{This refers to the challenge of integrating data with differing meanings or interpretations. 
In agile development, where teams often work with diverse tools and datasets, conflicting data semantics can lead to misalignment in decision-making and hinder collaboration. Resolving these inconsistencies is essential to ensure accurate data integration and good understanding within teams.}
Barcellos \cite{barcellos2020towards} discussed the difficulty of integrating data from different agents and tools due to semantic heterogeneity. Rix et al. \cite{rix2016agile} discussed the difficulty of integrating and making sense of diverse data sources with varying formats and meanings within the high-pressure die casting (HPDC) manufacturing process.  

\textbf{To resolve the semantic heterogeneity challenges}, Rosenkranz et al. \cite{rosenkranz2017supporting} proposed a communication-centric approach that combines agile software development practices with communication theory. This approach includes three core artifacts: a detailed template for data field specifications, a procedure model for iterative development, and agile methods to facilitate continuous stakeholder communication. These components work together to enhance stakeholder comprehension and ensure consistent data usage across the organization. The solution was implemented and evaluated. Dos Santos Júnior et al. \cite{dos2021ontology} proposed developing an ontology-based approach known as \textit{Immigrant} as a solution. By mapping the diverse semantics of various applications onto a common ontology, \textit{Immigrant} seeks to resolve semantic conflicts and provide a cohesive and integrated view of the data. This enables the presentation of meaningful information through dashboards, facilitating data-driven decision-making in agile software organizations. The approach is currently in development, with plans for a proof of concept to validate its effectiveness. In addition to the data harmonization and interoperability challenge addressed by Abdallah and Fan \cite{abdallah2022towards} suggested the development of ontologies discussed above to resolve data harmonization, interoperability, and semantic heterogeneity in heterogeneous MRO systems.

Barcellos \cite{barcellos2020towards} proposed using ontologies to establish a common conceptualization across different tools and data sources. This approach helps to reduce semantic conflicts and enables correct data integration  by providing a shared conceptual framework to align and
map the concepts used by various agents and tools, thus facilitating understanding and communication. They developed and used domain ontologies as reference models to integrate tools that support Continuous Software Engineering (CSE) processes. They started by developing and integrating a Scrum Reference Ontology with ontologies from the Software Engineering Ontology Network (SEON), covering requirements and project management aspects. This ontology was then applied as a basis to integrate tools such as Clockify and Azure DevOps used by development teams in the software unit of a Brazilian government agency. As a result, data from different tools were integrated and displayed in dashboards, providing useful information for managers to make decisions. Rix et al. \cite{rix2016agile} proposed a solution that implemented a modular and agile information processing framework. This framework utilizes annotation services, which add metadata to data sources, and a Master Data Management (MDM) repository, a centralized database that stores and manages master data. It enables real-time data streaming for data mining analyses and visualization of results. The framework ensures that data from various sources are annotated with metadata, facilitating pervasive traceability and simplifying information aggregation. Additionally, using dynamic Object-Relational Mapping (ORM) frameworks, such as Hibernate, in conjunction with open interfaces like OPC-UA (Open Platform Communications Unified Architecture), addresses semantic heterogeneity. This is accomplished by enabling the labeling and exchange of information in a unified manner. The solution was implemented and evaluated at the Audi testing foundry.\\

\noindent \textbf{Data Transformation and Extraction:}
\textcolor{black}{Data transformation involves converting data from diverse formats and sources into a usable form for analysis and decision-making. Agile teams often need to handle real-time data, requiring efficient processes to clean, format, and integrate information. Ineffective transformation and extraction can delay insights, disrupt workflows, and impede the agile principle of continuous delivery.}
Hofer et al. \cite{hofer2020new} the challenge of extracting and refining data from semi-structured, mixed-quality crowd-sourced data such as Wikipedia and Wikidata. Dursun et al. \cite{dursun2014workflow} emphasized the need for a unified view of data from various sources and to integrate data from heterogeneous stores into a single, coherent data store for analytics and visualization. Kannan et al. \cite{kannan2017rapid} discussed the importance of an efficient ETL process for extracting data from Electronic Health Records (EHR) into an Enterprise Data Warehouse (EDW). Dharmapal et al. \cite{dharmapal2016big} highlighted the complexity of integrating diverse data types without complex, time-consuming IT engineering efforts and suggested that modern analytic processing tools should transition away from traditional, retrospective BI tools and platforms towards more progressive analytic platforms.

\textbf{To address the data transformation and extraction challenges},  Hofer et al. \cite{hofer2020new} proposed a solution for improving the DBpedia Information Extraction Framework (DIEF) through the implementation of advanced ETL procedures. This approach included a systematic and test-driven method for handling data and code-related issues using Linked Data, which enhanced traceability and issue management. Additionally, the solution introduced two key technical enhancements: explicitly associating data artifacts with the corresponding code and implementing a library for continuous testing and validation of the data extraction process. The solution was implemented and evaluated. Dursun et al. \cite{dursun2014workflow} proposed implementing a data visualization layer to simplify user access to data, thereby abstracting the complexities of different storage structures and technologies. In addition, it suggested the establishment of a data federation layer to integrate data from various autonomous stores into a single cohesive data store for users. Furthermore, the proposed solution included an extensive modeling process. This process encompasses defining the goal to comprehend the business value, preparing the data, selecting the type of data-driven analytics, choosing suitable tools such as data mining, machine learning, optimization, and fuzzy inference systems, selecting tasks and algorithms from each tool, building the model, validating the model, and ultimately deploying the model. Kannan et al. \cite{kannan2017rapid} proposed developing a replicable ETL process that allowed the extraction of data from the EHR into the EDW using standard EHR fields and a core set of EHR structures for capturing custom data. This process supported the parallel development of EHR-based registries and associated data collection tools. The solution was implemented and evaluated. Dharmapal et al. \cite{dharmapal2016big} suggested a shift from conventional, retrospective business intelligence (BI) tools towards more advanced analytic platforms that support seamless integration with diverse data sources, including external ones. It further recommended adopting an agile methodology to adapt flexibly to changes in data sources, methodologies, algorithms, or tools to achieve business goals.\\

\noindent \textbf{Managing Data Integration:}
\textcolor{black}{This requires combining and aligning data from multiple systems into a unified dataset. Frequent updates and evolving requirements in agile environments add complexity to this process. Without an effective integration process in place, teams may struggle to generate cross-organizational insights, leading to inefficiencies and limited data-driven decision-making.}
Vestues et al. \cite{vestues2022agile} discussed the difficulty of managing data across various silos within the organization and supporting analytical solutions that require cross-organizational insights. Little et al. \cite{little2004adaptive} mentioned that a key business value of Landmark Graphics (a leading supplier of software and services) application suite arises from the integration of these products through a standard data model with over 800 tables, 12  entities, and 90 data types. This complexity presents a challenge in managing and integrating data across different products. Chen et al. \cite{chen2016agile} discussed the challenge of effectively managing internal and external data integration in big data projects, including data from existing Enterprise Data Warehouses (EDW) and new NoSQL systems.

\textbf{To address the challenges related to managing data integration},  Vestues et al. \cite{vestues2022agile} proposed the adoption of a ``data mesh'' approach. This approach comprises four core principles: domain-oriented decentralized data ownership and architecture, data as a product, self-serve data infrastructure, and federated computational governance. The data mesh model advocates for a shift from centralized data management to a distributed model where domain teams have increased ownership of the data produced by their applications, aiming to improve coordination and enable more agile and automated approaches to data analytics. Vestues et al. \cite{vestues2022agile} focused on reporting findings from a case study of a public sector organization in Norway that has begun the transition from centralized to distributed data management. Little et al. \cite{little2004adaptive} presented that Landmark Graphics adopted the Microsoft Solutions Framework (MSF), a milestone-based iterative development framework, to standardize their development process across different product teams. To manage the integration of diverse data types effectively of integrated products and ensure consistency in the integration efforts, which was crucial given the complexity of their common data model. Chen et al. \cite{chen2016agile} adopted an architecture-centric approach, which strongly emphasizes designing and developing a system's architecture as a crucial element of the development process. This approach entails the creation of a well-defined architecture, which acts as a blackprint for the system's development. It guides the selection of technologies, allocation of resources, and risk management throughout the development life cycle. This approach complements agile's modular design, a software development approach that emphasizes breaking down a project into smaller, more manageable modules or components to effectively manage the integration of diverse data types. The approach includes using data lakes and cloud storage to handle various data types, supporting agile's requirement for flexible data structures and continuous delivery. The new methodology,  AABA (Architecture-centric Agile Big Data Analytics), was implemented and validated through multiple case studies encompassing 10 big data analytics projects.\\

\noindent \textbf{Diverse and Decentralized Data Sources:}
\textcolor{black}{ Having diverse data sources poses a significant challenge for agile teams operating in distributed environments. Data is often fragmented across systems, teams, and geographic locations, making it difficult to consolidate and analyze effectively. This decentralization can lead to duplication of effort, delays in accessing key insights, and inconsistencies in data quality, all of which undermine the agility and responsiveness of teams.}
Upender \cite{upender2005staying} discussed the complexity involved in integrating patient data and transitioning to a centralized data management system within a diverse and decentralized user environment. Harper and Dagnino \cite{harper2014agile} highlighted the need to merge and clean data sets from different sources to enable advanced data analytics. The challenge is initiating and sustaining processes while ensuring data exchange and requests between components. The proposed solution is to distribute work requests across computing resources. Chhillar and Sharma \cite{chhillar2019act} discussed the challenge of consolidating reports from various sources, such as development, testing, and bug-tracking tools, into a unified data analytics tool. This challenge pertains to seamlessly integrating diverse data sources and formats to generate real-time metrics and reports for software quality analysis.

\textbf{To address the diverse and decentralized data sources challenges}, Harper and Dagnino \cite{harper2014agile} proposed developing a standard product line architecture with built-in capabilities for data cleansing and integration, designed to serve a wide range of domains and reduce startup costs for new analytics applications. The solution was implemented as part of an agile evolutionary approach to build the advanced analytics product line architecture for various industrial application scenarios at ABB (ASEA Brown Boveri), a multinational corporation. Chhillar and Sharma \cite{chhillar2019act} proposed an Automated Continuous Quality (ACQ) Metrics dashboard. This dashboard integrates data from various sources, such as functional, performance, security testing, Continuous Integration and Deployment (CI/CD) build details, and development reports into a data analytics tool. This integration facilitates the auto-generation of quality metrics, enabling real-time tracking and graphical representation of metrics. Chhillar and Sharma \cite{chhillar2019act} also noted that further endeavors are necessary for its implementation. Upender \cite{upender2005staying} proposed a solution, discussed in Section \ref{sub-DCCS}.

\subsubsection{Data Collection Challenges and Solutions}
\label{sub-DCCS}
We have identified multiple challenges associated with data collection in agile development. The studies reviewed below discuss these challenges and their proposed solutions are discussed in the studies reviewed below. 
A summary of these challenges and their corresponding solutions is presented in Table \ref{tab:tab-DataCollectionChallengesandSolutions}.\\

\begin{table}[]
\caption{\textcolor{black}{Summary of Data Collection Challenges and Solutions identified by the SLR.}}
\centering
\resizebox{\textwidth}{!}{%
\begin{tabular}{llll}
\toprule
\textbf{Challenge}     & \textbf{Solution}                                                           & \textbf{Solution Status}                                             & \textbf{Ref.}    
\\
\midrule
\multirow{3}{*}{Capturing Diverse Data}                 & User-centered design strategies                                           & Implemented                                                          & \cite{pater2018advancing}        \\
                                       & Continuous Integration tools& Proposed                                                             & \cite{matthies2019towards}       \\
                                       & Analytics-Driven Testing (ADT)                                                        & Proposed                                                             & \cite{batarseh2018predicting}    \\
\midrule
\multirow{5}{*}{\begin{tabular}[l]{@{}l@{}} Data Collection \\ Methods\end{tabular}}       & Automated Testing Dashboards                          & \multirow{2}{*}{\begin{tabular}[l]{@{}l@{}}Partially\\ Implemented \end{tabular}}                                 & \cite{chhillar2019act}           \\
&\\
                                       & Toolchain automation                                    & Proposed                                                             & \cite{lehtonen2017visualizations}\\
                                       & Human-centred agile platform                                      & Proposed                                                             & \cite{lin2018towards}            \\
                                       & Automated methods                                 & Proposed                                      & \cite{svensson2019unfulfilled}   \\
                                       & Centralized data management                                & Proposed                                             & \cite{upender2005staying}        \\
\midrule
\multirow{3}{*}{\begin{tabular}[l]{@{}l@{}} Data Sharing and \\ Collaboration\end{tabular}}         & Diagnostic model & Proposed                                                             & \cite{fabijan2016lack}           \\
                                       & DDSE methodologies                                          & Proposed                                                             & \cite{kaur2020dialogue}                   \\
                                       &Include a legal advisor for compliance                                                & Proposed                                                             & \cite{barbala2023data}           \\
\midrule
\multirow{4}{*}{\begin{tabular}[l]{@{}l@{}} Informative Data \\ Collection\end{tabular}}   & Ontology-based approach                                            & Proposed & \cite{dos2021ontology}           \\
                                       & Integration project data into Retrospective                            & Proposed                                                             & \cite{matthies2020playing}       \\
                                       & Q-Rapids tool                                             & Implemented                                                          & \cite{martinez2019continuously}  \\
                                       & Data analyst collaboration                  & Proposed                                                             & \cite{dharmapal2016big}          \\
\midrule
\multirow{4}{*}{\begin{tabular}[l]{@{}l@{}} Comprehensive Data \\ Collection\end{tabular}}            & Creation of a generalized dataset& Proposed                                                             & \cite{das2015towards}            \\
                                       & Big data stack                                             & Proposed                                                             & \cite{huang2021leveraging}       \\
                                       & Project Management Information System                  & Proposed                                                             & \cite{fagarasan2023integrating}  \\
                                       & Automated data collection mechanisms                                     & Proposed                                                             & \cite{olsson2018challenges} \\    
\bottomrule
\end{tabular}
}
\label{tab:tab-DataCollectionChallengesandSolutions}
\end{table}

\noindent \textbf{Capturing Diverse Data:}
\textcolor{black}{This involves collecting information from various sources and formats to meet agile development needs. In agile projects, diverse data may originate from user feedback, software metrics, and operational environments, each with unique structures and characteristics. The challenge lies in gathering this data systematically and ensuring its relevance for decision-making.}
Pater et al. \cite{pater2018advancing} focused on capturing data for personalized healthcare and human services while addressing barriers to widespread adoption and standardization of information systems within healthcare. Matthies and Hesse \cite{matthies2019towards} discussed the need for effective data collection and analysis from diverse sources, including software project data, customer feedback, and operational metrics. Furthermore, Batarseh and Gonzalez \cite{batarseh2018predicting} highlighted the issue of data is often not collected in a structured and organized manner, making it difficult to analyze and use for decision-making.

\textbf{To address capturing diverse data challenges}, Pater et al. \cite{pater2018advancing} proposed a solution that involves leveraging a variety of user-centered design strategies, such as participatory design, empathic design, and design thinking, to guide the data collection methods. This approach maintains the user at the center of the process and continuously incorporates user-driven input into each phase of the agile development cycle. The solution was implemented to modernize the Army Community Service's information technology infrastructure. Matthies and Hesse \cite{matthies2019towards} highlighted tools such as Continuous Integration servers, static analysis tools, and other software development tools that provide valuable data points on the current status and health of the developed software project. They highlighted that combining data analysis and interpretation by agile teams can enable better-informed business and software development decisions. Batarseh and Gonzalez \cite{batarseh2018predicting} proposed the Analytics-Driven Testing (ADT) process. ADT involves a structured data collection and analysis approach within the agile software development lifecycle. During each sprint, data are systematically collected using the ADT tool. This includes recording all software failures, measuring the Mean Time Between Failures (MTBF), and collecting other relevant data such as software module, sprint number, and run time.\\

\noindent \textbf{Data Collection Methods:}
\textcolor{black}{Challenges related to data collection methods arise when agile teams rely on manual, error-prone, or inconsistent methods for gathering data. Agile environments demand real-time, automated, reliable data collection processes to maintain efficiency.}
Chhillar and Sharma \cite{chhillar2019act} discussed the difficulty in generating real-time software quality metrics and the need for actual data collection processes. In contrast, Lehtonen et al. \cite{lehtonen2017visualizations} discussed the manual nature of data collection from the issue management system, which can be time-consuming and susceptible to errors.  Upender \cite{upender2005staying} discussed the challenge involves creating a system that allows for the electronic capture of clinical research data, replacing the existing rudimentary tools such as Excel and Access that various independent research groups use. Lin et al. \cite{lin2018towards} discussed the difficulty of collecting data on software engineering activities, such as task allocation, collaboration, and mood stability, without disrupting the normal Agile Software Development (ASD) process. Practitioners surveyed by Svensson et al. \cite{svensson2019unfulfilled} expressed concerns regarding data availability, excessive quantity, and ambiguity regarding usability, relevance, and its connection to decision-making.

\textbf{To address the challenges related to the data collection method}, Chhillar and Sharma \cite{chhillar2019act} proposed an Automated Continuous Testing (ACT) TestBot and ACQ Metrics dashboard. The ACT TestBot automates the execution of tests and the monitoring of application logs, while the ACQ Metrics dashboard aggregates and visualizes the data, enabling the automatic generation of quality metrics reports. This integrated approach ensures the reliability of data and facilitates the real-time tracking of software quality. Chhillar and Sharma \cite{chhillar2019act} noted the implementation requires more effort. Lehtonen et al. \cite{lehtonen2017visualizations} recommended automating the data collection by implementing a toolchain that covers all steps from the initial data collection to visualization. This automated approach would enhance the accuracy and efficiency of data collection, thereby improving the quality of the visualizations used to analyze the software development process. Additionally, the development of an interactive visualization tool was suggested to recognize patterns and anomalies in the process, facilitating a deeper understanding and validation of the evolution towards a more agile process.

Lin et al. \cite{lin2018towards} presented a Human-centred Agile Software Engineering (HASE) platform. The HASE platform is an online Agile Project Management (APM) tool allowing unobtrusive data collection during the normal ASD process. It collects data on software engineering activities from the participants, such as task allocation, collaboration, and mood stability, without requiring additional effort from the participants. This approach ensures that the data collection process does not disrupt the regular workflow of the software engineering teams while still providing accurate and objective data for assessing software engineering skills. Svensson et al. \cite{svensson2019unfulfilled} underscored that future research should develop novel automated methods for collecting, analyzing, and visualizing data to enhance existing agile decision-making processes by associating pertinent data with specific scenarios. Upender \cite{upender2005staying} proposed a three-year project to build a centralized data management system that addresses the needs of electronically collecting and integrating data with a gradual adoption and adaptation of Scrum and XP (Extreme Programming) practices. They emphasized a "just in time design" approach, which meant they approached the design phase with a flexible mindset. They used an evolutionary design process to handle ambiguous and conflicting requirements. Initially, they built a basic version of the clinical application to ensure it had essential functions, leaving the implementation of more advanced features for future updates. At the time the study was written, they were two years into the project.\\

\noindent \textbf{Data Sharing and Collaboration:}
\textcolor{black}{This challenge concerns ensuring data is accessible and shared seamlessly across agile teams. Limited access to data 
and fragmented storage often hinder collaboration and transparency.}
Fabijan et al. \cite{fabijan2016lack} noted the lack of customer and product data sharing among agile team members. This leads to difficulties in accessing data collected by others in different development phases and fragmented collection and storage of data. Kaur et al. \cite{kaur2020dialogue} pinpointed concurrent access as a significant challenge, leading to unnecessary rework due to human error and inefficient communication methods. Barbala et al. \cite{barbala2023data} discussed the challenge of the uncertainty surrounding what data agile product teams are allowed to gather.

\textbf{To address the data sharing and collection challenges}, Fabijan et al. \cite{fabijan2016lack} developed a model that serves as a diagnostic tool to understand data collection and sharing practices within software development organizations. This model identifies the types of data collected, the parties responsible for collection, and the development phases in which the data is used. It also highlights critical handovers where data loss can occur. The model can provide valuable insights for companies deciding on actions to improve their data sharing and collection practices. This can aid in addressing the fragmented collection and storage of data, ultimately leading to better data analysis and utilization. Kaur et al. \cite{kaur2020dialogue} recommended using Data-Driven Systems Engineering (DDSE) methodologies that facilitate concurrent access, enabling real-time collaboration and information exchange. DDSE tools manage engineering data during the implementation phase, provide version control, make it available collaboratively, and ensure full traceability for the entire engineering team. Barbala et al. \cite{barbala2023data} presented that including a legal advisor in the team helped assess the sensitivity of data when collecting, sharing, and storing it and provided guidance on compliance with data protection regulations.\\

\noindent \textbf{Informative Data Collection:}
\textcolor{black}{This challenge refers to the difficulties agile teams face in gathering meaningful and actionable data for decision-making. While data is abundant, its quality and relevance often fall short of agile requirements.}
Dos Santos Júnior et al. \cite{dos2021ontology} delved into difficulty in accessing, integrating, analyzing, and viewing data handled by heterogeneous applications, which often adopt different semantics (the same information item is given divergent interpretations), posing a barrier to integrated data usage. Matthies \cite{matthies2020playing} discussed the challenge of relying on subjective data gathered from team members' perceptions and experiences during Retrospective meetings. While readily available and relevant to team satisfaction, this subjective data may not provide a comprehensive or objective view of the team's performance and areas for improvement. Martínez-Fernández et al. \cite{martinez2019continuously} addressed the data collection challenge of obtaining informative user data. Dharmapal and Sikamani \cite{dharmapal2016big} underscored the significance of collecting data from suitable sources and filtering out unwanted data in the development phase.

\textbf{To address the informative data collection challenges}, Dos Santos Júnior et al. \cite{dos2021ontology} proposed a solution ``\textit{Development of ontology-based approach Immigrant}'' discussed in Section \ref{sub-DICS}. This solution resolved the semantic heterogeneity data integration challenge and assisted in collecting informative data that can be used for data-driven decision-making in agile software development. Matthies \cite{matthies2020playing} recommended integrating development project data and insights from software repository mining into Retrospective agendas. By employing project data from the last iteration in Retrospective meetings, teams can gain additional insights based on their team-specific data. This approach aims to provide a more thorough overview of the team's state, including facts and feelings, which can lead to better results in Retrospectives. Project data measurements can help track progress on common agile and software engineering challenges, enabling teams to identify and tackle improvement actions based on empirical project evidence rather than solely on anecdotal experiences. Martínez-Fernández et al. \cite{martinez2019continuously} proposed using the Q-Rapids tool for acquiring valuable information for users. The Q-Rapids tool is designed to gather diverse data related to software system development and usage. These data are then structured within a Quality Model (QM) to analyze aggregated quality-related strategic indicators. These indicators are made available to decision-makers through a multi-dimensional and navigational dashboard employed during ASD events like sprint planning or daily stand-up meetings. The Q-Rapids solution was implemented and used in a real-world setting. Martínez-Fernández et al. \cite{martinez2019continuously} described a case study across four companies involving 26 practitioners to investigate the integration of a Quality Model within the Q-Rapids software analytics tool. Dharmapal and Sikamani \cite{dharmapal2016big} proposed utilizing a combination of agile methodology principles and the expertise of a Data Analyst. During the planning stage, the Product Owner and the Data Analyst collaborate to select suitable Big Data Analytics tools and strategize the data source collection. This approach allows for incremental data collection in response to specific needs. In the development phase, the Data Analyst is pivotal in analyzing data gathered from various sources. They integrate data from diverse origins into a unified dataset, merge information based on shared attributes, and present it in a consolidated format. An essential aspect of their role is removing irrelevant data, employing an elimination technique that saves time and resources by excluding unnecessary data from processing.\\

\noindent \textbf{Comprehensive Data Collection:}
\textcolor{black}{This involves collecting a wide range of data from various sources to support agile decision-making. The challenge lies in processing and integrating diverse datasets, especially in environments with large-scale or unstructured data.}
Das et al. \cite{das2015towards} delved into the challenges associated with collecting and storing substantial amounts of data for specific research purposes, mainly when the data is non-traditional or the dataset is immense. Huang et al. \cite{huang2021leveraging} focused on efficiently ingesting and governing large volumes of heterogeneous data from various sources in various formats. Fagarasan et al. \cite{fagarasan2023integrating} stressed the importance of systematic and comprehensive data collection, beginning at the lowest level of project management. Olsson \cite{olsson2018challenges} discussed that traditional data collection methods, such as surveys and interviews, are insufficient for continuous experimentation.

\textbf{To address the comprehensive data collection challenges}, Das et al. \cite{das2015towards} advocated for a two-fold approach: first, the creation of a generalized dataset aimed at expanding data coverage (not volume) by incorporating additional attributes. This approach ensures that a single data extraction step can serve multiple projects, with adjustments made in subsequent workflows to cater to specific research task requirements. Second, they emphasized standardized data processing, which involves abstracting standard data procedures, such as noise reduction and language standardization, into reusable solutions and tools applicable across various data-driven research endeavors. It makes Datto systematically collect and analyze.  Huang et al. \cite{huang2021leveraging} The solution they proposed was to design a big data stack with scalable storage technologies and a flexible architecture. This stack included a data lake for raw data storage and integrated relational databases for real-time applications. Apache Spark was used as a distributed data computation engine to enable parallel processing for analytics tasks. The stack's design allowed for swift data analysis and agile development of data products, reducing the time required to answer analytical questions and develop applications. Fagarasan et al. \cite{fagarasan2023integrating} proposed using a market-available Project Management Information System (PMIS) to collect and analyze data systematically. This system enables the automated collection and reporting of key performance indicators, which are crucial for monitoring project performance and ensuring cost-effective execution. The PMIS is a foundation for organizing projects into a measurable portfolio, with performance metrics deeply embedded in the software development workflow. Olsson \cite{olsson2018challenges} proposed implementing mechanisms for automated data collection directly from the software in use, particularly in the post-deployment stage. This approach allows for the collection of real-time and actionable data that can be used for analysis to improve current products and inform the development of future ones, aligning with legal requirements for data utilization and user consent.


\subsubsection{Data Quality Challenges and Solutions}
\label{sub-DQCS}

We have identified multiple challenges associated with Data Quality in agile development. These challenges and their proposed solutions are discussed in the studies reviewed below. 
A summary of these challenges and their corresponding solutions is presented in Table \ref{tab:tab-DataQualityChallengesandSolutions}.\\

\begin{table}[]
\caption{\textcolor{black}{Summary of Data Quality Challenges and Solutions Identified by the SLR.}}
\centering
\adjustbox{width=\linewidth}{ 
\begin{tabular}{llll}
\toprule
\textbf{Data Quality Challenge}                                & \textbf{Solution}                                                       & \textbf{\begin{tabular}[c]{@{}l@{}}Implementation  \\ Status\end{tabular}} & \textbf{Ref.}                                      \\
\midrule
\multirow{2}{*}{\begin{tabular}[c]{@{}l@{}}Ensuring Data Accuracy and \\ Consistency Across Varied Sources\end{tabular}}

& Ref in Sec. \ref{sub-DICS}              & Implemented              & \cite{rosenkranz2017supporting}   \\
                                                               & Ref in Sec. \ref{sub-DICS}  & Implemented              & \cite{harper2014agile}            \\
                                                               & Ref in Sec. \ref{sub-DICS}           & Implemented              & \cite{little2004adaptive}         \\
                                                               & Ref in Sec. \ref{sub-DCCS}              & Implemented              & \cite{pater2018advancing}         \\
                                                               & Ref in Sec. \ref{sub-DICS}      & Implemented              & \cite{kannan2017rapid}            \\
\midrule
Missing Quality Data                                           &\multirow{2}{*}{\begin{tabular}[c]{@{}l@{}} Utilizing a quality\\-aware models\end{tabular}} & Implemented              & \cite{franch2019quality}          \\
& \\
                                                               & Ref in Sec. \ref{sub-DCCS}              & Proposed                 & \cite{lehtonen2017visualizations} \\
                                                               & Ref in Sec. \ref{sub-DCCS}                   & Proposed                 & \cite{martinez2019continuously}   \\
\midrule
\multirow{2}{*}{\begin{tabular}[c]{@{}l@{}}Inadequate Data Quality\\  Management\end{tabular}}
                             & \multirow{2}{*}{\begin{tabular}[c]{@{}l@{}} Using test-driven \\approaches\end{tabular}}                                 & Proposed                 & \cite{ambler2008gets}             \\
                              &\\
                                                               & Ref in Sec. \ref{sub-DICS}      & Implemented              & \cite{hofer2020new}               \\
                                                               & Ref in Sec. \ref{sub-DICS} & Implemented              & \cite{svensson2019unfulfilled}    \\
\midrule
Data Quality Standardization                                   & Ref in Sec. \ref{sub-DCCS}         & Implemented              & \cite{spengler2020enabling}       \\
                                                               & Ref in Sec. \ref{sub-DCCS}              & Proposed                 & \cite{batarseh2018predicting}   \\
\bottomrule
\end{tabular}
}
\label{tab:tab-DataQualityChallengesandSolutions}
\end{table}

\noindent\textbf{Ensuring Data Accuracy and Consistency Across Varied Sources:}
\textcolor{black}{This challenge arises when data from multiple systems and sources must be aligned to ensure reliability and usability. In agile environments, where frequent updates and dynamic data integration are common, maintaining accuracy and consistency is critical to avoid errors and inefficiencies.}
Rosenkranz et al. \cite{rosenkranz2017supporting} highlighted the importance of data quality (defined in their study as completeness, ambiguity, meaningfulness, and correctness) in application systems and project performance in IT or Data Warehouse (DW) projects. The challenge lies in ensuring the accuracy and consistency of data from disparate sources. Harper and Dagnino \cite{harper2014agile} also emphasized the need to merge and clean datasets from different sources. The challenge here is ensuring the accuracy and consistency of data for analysis. Little et al. \cite{little2004adaptive} mentioned that the products are released regularly, with release cycles ranging from 3 to 18 months. This regularity poses a challenge in managing data and ensuring consistency across different product versions. Pater et al. \cite{pater2018advancing} underscored the necessity to eradicate data redundancy and business process fragmentation in the context of IT system modernization. The challenge lies in aligning the diverse IT needs of users with the broader organizational data requirements. Kannan et al. \cite{kannan2017rapid} highlighted the challenge of ensuring the completeness, accuracy, and internal consistency of the critical few EHR fields essential for registry inclusion and eCQM (electronic Clinical Quality Measure) calculation.

\textbf{To address the data accuracy and consistency challenges across varied sources:} The solutions are the same as those of the data integration and collection (as those challenges co-occur together). We discussed those solutions in detail in Sections \ref{sub-DICS} and  \ref{sub-DCCS}.\\

\noindent \textbf{Missing Quality Data:}
\textcolor{black}{This challenge pertains to the absence of complete, accurate, or usable data, which can hinder agile teams' ability to make informed decisions. The lack of quality data in agile development often stems from incomplete data collection processes or insufficient tools.}
Franch et al. \cite{franch2019quality} emphasized the issue when necessary quality data is unavailable (i.e., inaccessible, incomplete, inconsistent, and incorrect data) due to the absence of suitable data collection tools or processes or inaccessible data storage systems. For instance, SonarQube (mentioned as an existing tool that focuses on product quality or continuous integration) does not furnish raw data for specific base metrics. Lehtonen et al. \cite{lehtonen2017visualizations} highlighted potential data quality problems in software engineering data, including noise, outliers, low precision, missing values, coverage errors, and clones. An example given is the potential for inaccurate data due to developers forgetting to update the issue management system task state when development work starts or ends, leading to inaccurate timestamps. Martínez-Fernández et al. \cite{martinez2019continuously} discussed the need for transparency and clarity on the raw data used to compute factors and indicators in the software analytics tool (Q-Rapids tool). End-users like product owners may require a clear understanding of the raw data used to derive specific values and the associated decision-making processes.

\textbf{To address the missing data} Franch et al. \cite{franch2019quality} discussed the Q-Rapids project, mentioned as an initiative aimed at developing a quality-aware rapid software development methodology. The Q-Rapids project addressed data quality challenges by selecting and configuring tailored data collection tools aligned with business objectives, specifying precision requirements, implementing real-time checks to prevent errors, and integrating diverse data sources while transforming them for compatibility with specific analysis methods and tools. The solution was implemented, evaluated, and refined through three releases deployed by the four industry partners in the project. Note that the solutions suggested to resolve the remaining data quality challenges, as discussed by Lehtonen et al. \cite{lehtonen2017visualizations} and Martínez-Fernández et al. \cite{martinez2019continuously}, are the same as those of data collection (as they co-occur together). We discussed those solutions in detail in Section \ref{sub-DCCS}.\\

\noindent \textbf{Inadequate Data Quality Management:}
\textcolor{black}{This refers to the failure to implement effective processes for monitoring, maintaining, and improving data quality throughout agile projects. Agile workflows often emphasize rapid delivery, which can lead to delayed focus on data quality until critical phases.}
Ambler \cite{ambler2008gets} highlighted that data quality issues cost US organizations an estimated 600 billion dollars annually (according to Data Warehouse Institute), stemming from the use of traditional data management approaches to ensure data quality within organizations, according to the Data Warehouse Institute as indicated by a survey conducted by Ambler \cite{ambler2008gets}. Svensson et al. \cite{svensson2019unfulfilled} emphasized that the quality of the data and the processing techniques and tools used to handle it will impact the quality of decisions made using Data-Driven Decision Making (DDDM). Hofer et al. \cite{hofer2020new} discussed the data quality challenge of ensuring the "fitness for use" of data, which is often neglected or delayed in the software engineering process until the end-user evaluation phase, known as the ``\textit{point-of-truth}''. This delay impacts the cost-effectiveness of data quality management and contradicts agile principles that emphasize early and continuous delivery of valuable software.

\textbf{To address inadequate data quality management}, Ambler \cite{ambler2008gets} Suggested that utilizing agile methods, which typically involve more and earlier test-driven approach to development, results in better quality with more and early testing. Additionally, developers and data professionals should apply concrete, quality-focused techniques for evolutionary development. Note that the solutions suggested to resolve the remaining data quality challenges, as discussed by Svensson et al. \cite{svensson2019unfulfilled} and Hofer et al. \cite{hofer2020new}, are the same as those of data collection and data integration (as they co-occur together). We discussed those solutions in detail in the data collection Section \ref{sub-DCCS} and for data integration Section \ref{sub-DICS}.\\

\noindent \textbf{Data Quality Standardization:}
\textcolor{black}{This involves ensuring that data adheres to consistent formats and standards to support integration and analysis. Achieving standardization is critical in agile environments, where data is sourced from diverse systems and rapidly evolving requirements.}
Spengler et al. \cite{spengler2020enabling} mentioned the need for significant data restructuring and cleansing due to heterogeneous data to ensure suitability for proper integration into clinical data warehousing solutions. Batarseh and Gonzalez \cite{batarseh2018predicting} highlighted the need to enhance dataset quality, standardize variables, and eliminate unnecessary data. The challenge here involves dealing with poor-quality data and standardizing data for analysis.

\textbf{To address the data quality standardization challenges:} The solutions are the same as those of data integration and collection (as they co-occur together). We discussed those solutions in detail for data integration in Section \ref{sub-DICS} and for data collection in Section \ref{sub-DCCS}.

\subsubsection{Data Analysis Challenges and Solutions}
\label{sub-DACS}
We also identified multiple challenges associated with Data analysis in Agile development. These challenges and their proposed solutions are discussed in the studies reviewed below. 
A summary of these challenges and their corresponding solutions is presented in Table \ref{tab:tab-DataAnalysisChallengesandSolutions}.\\

\begin{table}[]
\caption{\textcolor{black}{Summary of Data Analysis Challenges and Solutions identified by the SLR.}}
\centering
\resizebox{\textwidth}{!}{%
\begin{tabular}{llll}
\toprule
\textbf{Data Analysis Challenge}               & \textbf{Data Analysis Solutions}                                                                                                      & \textbf{Solution Status} & \textbf{Ref.}                                      \\
\midrule
\multirow{2}{*}{\begin{tabular}[c]{@{}l@{}}Analyzing Large\\ and Complex Data \end{tabular}}&Ref in Sec. \ref{sub-DICS}                                                             & Proposed                 & \cite{chen2016agile}              \\
                                                 &Ref in Sec. \ref{sub-DCCS}                                                           & Proposed                 & \cite{dharmapal2016big}           \\
                                                 &Ref in Sec. \ref{sub-DCCS}                                                         & Proposed                 & \cite{das2015towards}             \\
                                                 &Ref in Sec. \ref{sub-DCCS}                                                                     & Proposed                 & \cite{lin2018towards}             \\
                                                 & \multirow{2}{*}{\begin{tabular}[c]{@{}l@{}}Well-documented tools\\and developer training.\end{tabular}}                                   & Proposed                 & \cite{hamer2023students}          \\
                                                 & \\
\midrule
\multirow{2}{*}{\begin{tabular}[c]{@{}l@{}}Analyzing Semantic\\ Heterogeneity Data\end{tabular}}&Ref in Sec. \ref{sub-DICS}                                                                & Proposed                 & \cite{dos2021ontology}            \\
                                                 &Ref in Sec. \ref{sub-DICS}                                                                & Implemented              & \cite{barcellos2020towards}       \\
                                                 &Ref in Sec. \ref{sub-DICS}                                                                & Implemented              & \cite{rix2016agile}               \\
\midrule
\multirow{2}{*}{\begin{tabular}[c]{@{}l@{}}Efficient Data Analysis\\ and Visualization\end{tabular}}&Ref in Sec. \ref{sub-DICS}                                                        & Proposed                 & \cite{dursun2014workflow}         \\
                                                 &Ref in Sec. \ref{sub-DCCS}                                                                     & Proposed                 & \cite{svensson2019unfulfilled}    \\
                                                 &Ref in Sec. \ref{sub-DCCS}                                                                     & Proposed                 & \cite{lehtonen2017visualizations} \\
                                                 &Ref in Sec. \ref{sub-DCCS}                                                            & Proposed                 & \cite{fabijan2016lack}            \\
\midrule
\multirow{2}{*}{\begin{tabular}[c]{@{}l@{}}Real-Time Data Analytics\\ and Decision Making \end{tabular}}&\multirow{2}{*}{\begin{tabular}[c]{@{}l@{}}Ref in Sec. \ref{sub-DICS}\\and Sec. \ref{sub-DCCS}\end{tabular}} & Proposed                 & \cite{chhillar2019act}            \\
&\\
                                                 &Ref in Sec. \ref{sub-DCCS}                                                         & Proposed                 & \cite{olsson2018challenges}       \\
\midrule
\multirow{2}{*}{\begin{tabular}[c]{@{}l@{}}Selection of Appropriate \\ Analytical Techniques\end{tabular}} &Ref in Sec. \ref{sub-DICS}                                                    & Implemented              & \cite{harper2014agile}            \\
                                                 &Ref in Sec. \ref{sub-DCCS}                                                                & Proposed                 & \cite{matthies2019towards}        \\
                                                 &Ref in Sec. \ref{sub-DCCS}                                                           & Proposed                 & \cite{matthies2020playing}        \\
                                                 &Ref in Sec. \ref{sub-DCCS}                                                         & Proposed                 & \cite{fagarasan2023integrating}   \\
                                                 &Ref in Sec. \ref{sub-DCCS}                                                       & Proposed                 & \cite{batarseh2018predicting}    
\\
\bottomrule
\end{tabular}%
}
\label{tab:tab-DataAnalysisChallengesandSolutions}
\end{table}

\noindent \textbf{Analyzing Large and Complex Data:}
\textcolor{black}{This challenge arises from the need to process and derive insights from massive, intricate datasets within agile environments. Agile teams often handle diverse data sources, requiring advanced tools and techniques to ensure timely and actionable analysis.}
Chen et al. \cite{chen2016agile} delved into the challenge of developing an extensive data system that effectively supports advanced analytics by addressing the 5Vs of big data (Volume, Velocity, Variety, Veracity, and Value) and facilitating effective collaboration between data scientists and software engineers to maximize the value from big data. Dharmapal and Sikamani \cite{dharmapal2016big} underscored the challenge of managing and processing the increased volume of unstructured data, which requires more complex enhancements in data management and analysis techniques. Das et al. \cite{das2015towards} discussed the challenge of analyzing extensive volumes of data, particularly when the data is non-traditional or when the dataset is massive. Traditional analysis techniques often fall short due to these challenges. Lin et al. \cite{lin2018towards} discussed the challenge of the high dimensionality of the datasets, which included detailed interactions and decisions made by the participants (students in the study). This complexity made it difficult to identify which features (which could include metrics such as the number of collaborators per task, the frequency of task updates, mood stability, and other behavioral indicators) or a combination of features could accurately predict certain behaviors of interest. Hamer et al. \cite{hamer2023students} highlighted that interpreting the data provided by data-driven tools such as Git, Jira, and SonarQube can pose a significant barrier to beginning the effective use of these tools due to the large volume and complexity of the data generated. Furthermore, analyzing this data necessitates training to utilize the information correctly.

\textbf{To address the analyzing large and complex data challenges:} Tsolutionsions are the same as those of the data integration and collection (as they co-occur together). We discussed those solutions in detail for data integration in Section \ref{sub-DICS} and for data collection in Section \ref{sub-DCCS}. Additionally, Lin et al. \cite{lin2018towards} solution is to employ the application of Exploratory Data Analysis (EDA) to analyze the collected data. EDA is often used to summarize the main characteristics of data sets using visual methods. It aims to understand what can be learned from the data beyond the formal modeling or hypothesis-testing tasks. Furthermore, Hamer et al. \cite{hamer2023students} solution recommends that these tools be well-documented and provide tutorials to facilitate easier adoption and usage. Developers should be trained to comprehend and leverage the provided information effectively. Future work could involve creating simplified visualizations that enhance information understanding. This proposed solution has not been implemented.\\

\noindent \textbf{Analyzing Semantic Heterogeneity Data:}
\textcolor{black}{This challenge involves addressing inconsistencies in the meaning and interpretation of data across diverse systems and sources. Semantic heterogeneity complicates data integration and analysis in agile environments, where teams often rely on disparate tools and data formats.}
Dos Santos J{'u}nior et al. \cite{dos2021ontology} highlighted the difficulties organizations encounter when accessing, integrating, analyzing, and viewing data managed by disparate applications. Barcellos \cite{barcellos2020towards} discussed the need for continuous software measurement, which involves collecting and analyzing data to provide useful information for daily activities and decision-making in software development. Rix et al. \cite{rix2016agile} addressed the challenge of analyzing real-time data streams using data mining methods.

\textbf{To address the analyzing semantic heterogeneity data challenges:} The solutions are the same as those of the data integration (as they co-occur together). We discussed those solutions in detail for data integration in Section \ref{sub-DICS}.\\

\noindent \textbf{Efficient Data Analysis and Visualization:}
\textcolor{black}{This refers to the challenge of transforming raw data into meaningful insights through streamlined analysis and clear visual representation. Agile projects demand rapid and effective visualization tools to aid in decision-making.}
Dursun et al. \cite{dursun2014workflow} discussed the challenges of efficiently analyzing and visualizing vast and diverse oil and gas data and developing intelligent data-driven analytics software to optimize decision-making and minimize. Svensson et al. \cite{svensson2019unfulfilled} addressed data analysis as a crucial step in data-driven decision-making (DDDM). The challenge lies in the fact that the quality of decisions is directly proportional to the quality of processing techniques and tools. In other words, ineffective or incorrect data analysis can lead to suboptimal or erroneous decisions. Lehtonen et al. \cite{lehtonen2017visualizations} explored the utilization of visualizations for data analysis. The challenge lies in the potential complexity of these visualizations, which may hinder understanding. Fabijan et al. \cite{fabijan2016lack} highlighted that companies are not capitalizing fully on the data they amass. This is attributed to challenges associated with efficiently and meaningfully integrating and analyzing customer data.

\textbf{To address the efficient data analysis and visualization challenges:} The solutions are the same as those of the data integration and collection (as they co-occur together). We discussed those solutions in detail for data integration in Section \ref{sub-DICS} and for data collection in Section \ref{sub-DCCS}.\\

\noindent \textbf{Real-Time Data Analytics and Decision Making:}
\textcolor{black}{This challenge focuses on enabling agile teams to process and utilize real-time data to support adaptive decision-making. Agile development’s iterative nature demands analytics systems that provide immediate feedback and insights.}
Chhillar and Sharma \cite{chhillar2019act} discussed using data analytics tools for generating metric reports. These reports are vital for indicating the overall health of the build post-deployment. The challenge here is the requirement for real-time data analytics to support decision-making processes. Olsson \cite{olsson2018challenges} discussed the difficulty in handling and analyzing the post-deployment data collected from embedded systems.

\textbf{To address the real-time data analytics and decision-making challenges:} The solutions are the same as those of the data integration and collection (as they co-occur together). We discussed those solutions in detail for data integration in Section \ref{sub-DICS} and for data collection in Section \ref{sub-DCCS}.\\

\noindent \textbf{Selection of Appropriate Analytical Techniques:}
\textcolor{black}{This involves choosing the right tools, methods, and frameworks to analyze data effectively within agile contexts. With diverse data types and frequent changes, selecting appropriate techniques becomes critical to ensuring relevance and accuracy.} Harper and Dagnino \cite{harper2014agile} discussed the need to specify analysis strategy, process substantial amounts of data, and complete analyses promptly. The challenge lies in fulfilling these expectations while managing the complexity inherent in advanced analytics applications. Matthies and Hesse \cite{matthies2019towards} discussed data generation during software development processes such as work item descriptions, documentation, and version control information. The challenge lies in efficiently managing and analyzing this data. Matthies \cite{matthies2020playing} discussed using various metrics designed for agile practices or tools that developers are already familiar with, such as the git command line, for project data analysis. The challenge is that improvement actions are often based on anecdotal evidence and experiences rather than empirical project evidence. Fagarasan et al. \cite{fagarasan2023integrating} discussed the necessity for effective data analysis to extract meaningful insights from collected data. The challenge lies in interpreting, analyzing, and converting this information into actionable insights. Batarseh and Gonzalez \cite{batarseh2018predicting} underscored the need to define variables for model development and investigate different models. The challenge here lies in selecting appropriate variables and models for analysis.

\textbf{To address the selection of appropriate analytical techniques challenges:} The solutions are the same as those of the data integration and collection (as they co-occur together). We discussed those solutions in detail in Sections \ref{sub-DICS} (data integration) and \ref{sub-DCCS} (data collection).



\subsection{Summary of Systematic Literature Review}

\begin{sidewaysfigure}
    \centering
    \includegraphics[width=0.75\linewidth]{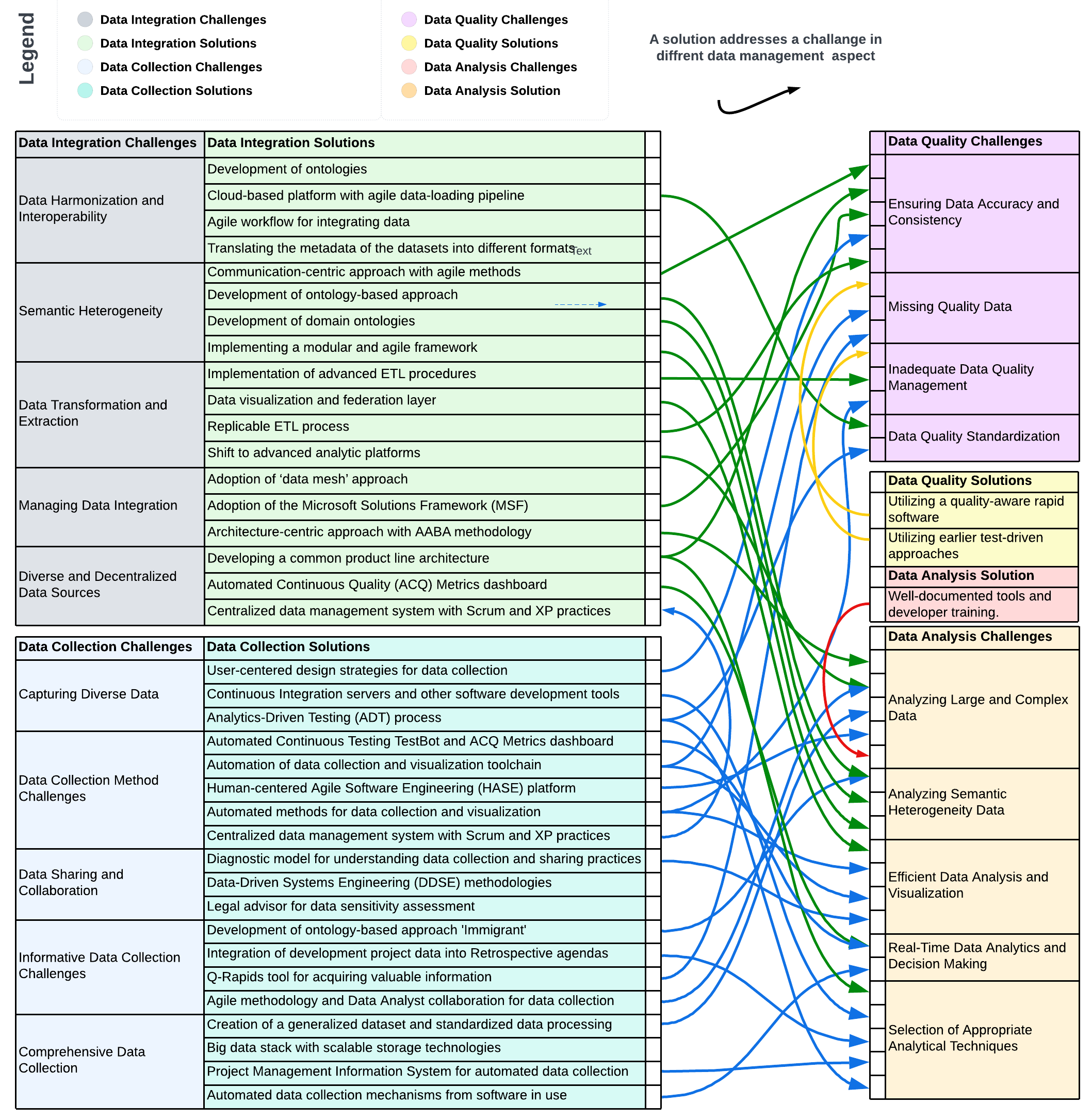}
    \caption{Summary: Data Management Challenges and Solutions from the SLR}
    \label{fig:DMCSSumary}
\end{sidewaysfigure}

Our SLR identified key data management challenges and solutions in agile software development. The review identified key data management aspects (with insights such as the overlapping between these aspects), including data integration, data quality, data collection, and data analysis. The SLR also revealed several solutions discussed to overcome those challenges.
Figure \ref{fig:DMCSSumary} summarizes the challenges and links to solutions.

To address the challenges identified in each aspect, we found that the proposed or implemented solutions vary across studies, depending on the nature of the systems and the maturity of the processes followed. We also note that most solutions are mainly proposed with no empirical evaluation. 


To further explore these challenges and the solutions proposed in the literature, we designed a practitioner survey in Section~\ref{sec:survey}. This survey aims to capture the practical experiences and insights from agile software development practitioners in various industries, complementing academic perspectives with specific project practices. By aligning the survey questions with the challenges and solutions identified in the SLR, we ensure that the survey not only confirms the relevance of these challenges in practice but also uncovers additional insights that may not be covered in the existing literature. This integrated approach allows us to provide a more holistic understanding of data management in agile software development, bridging the gap between theory and practice. 

\section{Practitioner Survey on Data Management}
\label{sec:survey}
Now that we have identified data management challenges and solutions from the SLR in Section \ref{sec:slr}, we used those findings to design our practitioner survey. Based on the challenges and solutions reported in the literature, the survey aims to better understand what practitioners identify as challenges (RQ1) and solutions (RQ2) and whether those challenges and solutions are shared among practitioners. Furthermore, feedback from practitioners across different industries can enhance the relevance of the insights from the SLR for the practical application of our results and for guiding future research directions.

\subsection{Survey Method}
\label{sec:methodology-surv}
 We have followed guidelines for conducting surveys as proposed by Punter et al. \cite{punter2003conducting} and Lin{\aa}ker et al. \cite{linaaker2015guidelines} in designing and conducting this survey. We discuss our survey design below.
 
\subsubsection{Survey Design}
\label{sec:survey:design}
The SLR informs the survey by guiding the development of specific sections based on data management aspects, ensuring comprehensive coverage of pertinent challenges (RQ1) and solutions (RQ2). For example, the SLR highlights challenges like data harmonization and semantic heterogeneity in data integration, leading to survey questions on specific integration challenges and solutions. Regarding data collection, the SLR identifies challenges in capturing diverse data, which is reflected in survey questions about data collection challenges and strategies. For data quality, the SLR emphasizes the need to ensure data accuracy and consistency, prompting survey questions on quality challenges and maintenance methods. The SLR findings on difficulties in analyzing complex data inform survey questions about analysis challenges and reliable techniques.
The SLR also identifies gaps in the literature, and the survey addresses them by asking for detailed practitioner experiences and solutions. For example, it directly asks the practitioners about the impact of these challenges on their role and the project delivery process.\\

\noindent\textbf{Target population and the sampling strategy:}
Our target population includes software development practitioners across various roles and experience levels in agile development projects. This encompasses senior developers, team leads, scrum masters, product owners, analysts, project managers, quality managers, DevOps engineers, data engineers, and solution architects. We employed a purposive sampling strategy \cite{patton2014qualitative}, aimed at selecting individuals experienced in agile software development and with practical insights into data management challenges and solutions. This approach ensures that the data collected is rich in detail and relevant to our research questions. We circulated the online survey link to practitioners through personal connections and local industrial networks. We aimed to distribute the survey to a broad audience, specifically those strongly interested in data management. The survey was distributed internationally to agile practitioners through direct contact at different organizations, through LinkedIn, and emails. The survey was distributed in March 2024, and the response period was open until June 2024. We closely monitored responses to ensure the progress of the data collection process. To boost the response rate, we sent follow-up reminders to the contacts list to encourage participation.\\

\noindent\textbf{Questionnaire development:}
The survey instrument (questionnaire) is divided into the following sections:

\begin{enumerate}[nosep] 
\item  \textit{Purpose and consent:} Participants are given information on the survey's objective and consent information (this was mandatory to answer). 
\item  \textit{Introduction:} This section explains the relevance of data management in agile projects. 
\item \textit{Background: } This section gathers demographics including age, gender, professional seniority, years of experience, educational background and role in agile projects, industry sector (pre-defined sectors with an option to add more), and professional information from participants. Additionally, we asked a mandatory question: ``\textit{Have you worked, or are you currently working on a project that follows agile software development practices?}''; if the answer was 'Yes,' the participant could continue answering the subsequent questions. 
\item  \textit{Instruction:} This section instructs the participants to base their answers on agile projects in which they were heavily involved. 
\item \textit{Data management challenges and solutions:} This section collects answers related to (data integration, collection, quality, and analysis) challenges and solutions based on the findings from the SLR (see Section \ref{sec:slr}), but also allowed participants to comment on their answers in an open-ended field. We employed a mix of multiple-choice and open-ended questions for each data management aspect section. For example, in the data integration section, we ask open-ended questions such as ``\textit{What data integration challenges have you encountered in your projects?}'', followed by multiple-choice to ask the same question with giving choices delivered from the SLR (see Appendix Table \ref{tab:AppDICH}). We employed the same style of questions (open-ended followed by multiple-choice) for each data management aspect section.
\item  \textit{Recommendations:} This section asks the participants to provide information about the impact of data management challenges on their respective roles, project delivery, and potential enhancements.

\end{enumerate}

The survey was approved by the university ethics committee. Participants could take the survey anonymously and choose not to provide information that could identify them. All data from the survey were securely stored online. 

Before deploying our survey, we conducted a pilot survey with three practitioners with more than ten years of experience in agile and data management (senior agile project manager, senior data analyst, and senior developer). The pilot aimed to verify the questions, scope, and estimated completion time. We used the feedback from the pilot to refine the questionnaire structure, clarify the questions, and address any identified validity threats. For example, we rephrased options for clarity and made structural changes to the survey flow. We included additional questions based on the pilot survey results.

The survey questionnaire can be accessed from our replication package \cite{fawzy_2024_10597818}

\subsubsection{Data Analysis}
\label{SureveyDataAnalysis}
We conducted a quantitative and qualitative analysis of the survey data we obtained from all participants. 

\textbf{Quantitative Analysis:}
We first gathered statistics from the responses, including demographic and response frequencies. We then investigated the general trends and patterns in the collected data.

\textbf{Qualitative Analysis:}
We followed the guidelines of Braun and Clarke's thematic analysis \cite{braun2006using}, employing a combined approach. 

\begin{itemize}
\item \textit{Deductive}: Using open-ended questions provided a deductive framework by defining broad categories. For example, questions like ``\textit{What data integration challenges have you encountered in your projects?}'' and ``\textit{What solutions have you employed to address data quality challenges?}'' helped establish categories such as data integration challenges, data integration solutions, and data quality challenges.

\item \textit{Inductive}: Specific themes and sub-themes were identified inductively from the explicit data content within those categories (example provided in the steps below).

\item \textit{Semantic}: The analysis focused on the explicit, surface meanings of the participants' responses rather than underlying assumptions or implicit meanings. For example, when a participant stated, ``\textit{We faced issues with data consistency across different datasets}'', it was directly coded as a ``\textit{Data Consistency Issue}'' without interpreting deeper underlying causes.

\end{itemize}

We carried out the qualitative analysis using NVivo\footnote{ \url{https://lumivero.com/products/nvivo/}}. We explain the steps taken for this analysis below: 

\begin{itemize}
    \item \textit{Data Familiarizing}: we first familiarized ourselves with the data by reading over the survey responses. Iterative rereading of the responses helped to identify initial themes and codes. At each step, we took notes on possible codes and themes.
    \item \textit{Developing Initial Codes}: the first author created the initial codes for each open-ended question. The second author reviewed the developed codes to ensure consistency.
    \item \textit{Identifying Themes}: codes were categorized into potential themes. Those themes were examined for consistency and coherence. Each theme was given a descriptive name that appropriately reflected its content. We also identified sub-themes as needed. Two of the authors iteratively discussed and refined the identified themes. For instance, the codes ``\textit{Inconsistent Data Structures}'' and ``\textit{different Data Formats}'' were categorized under  ``\textit{Data Structure and Format Challenges}''. 
    \item \textit{Reviewing Themes}: to strengthen validity, the first author reviewed the themes while the second author confirmed them. We conducted two review phases:
    \begin{enumerate} 
         \item { \textbf{Level 1}: Reviewing the themes in light of participants' quotes under the theme:} We checked participants' quotes to ensure they accurately represented that theme.
       \item {\textbf{Level 2}: Reviewing the themes in light of the entire dataset:} We looked at the themes in the context of all the data. This meant checking how well the themes represented the entire set of responses and making adjustments if needed to ensure they accurately reflected the data and our RQs.
       \end{enumerate} 
    \item \textit{Comparison with SLR-Based Options}: we compared the identified themes from the open-ended question with the options in the multiple-choice question. This helped integrate the findings into the results section.
\end{itemize}

\subsection{Results}
We present our survey results below. We received 32 responses from practitioners. Before we present the results of our analysis of the survey data, we introduce demographic information about the participants. 

\subsubsection{Demographics}

\begin{sidewaysfigure}
    \centering
    \includegraphics[width=1\linewidth]{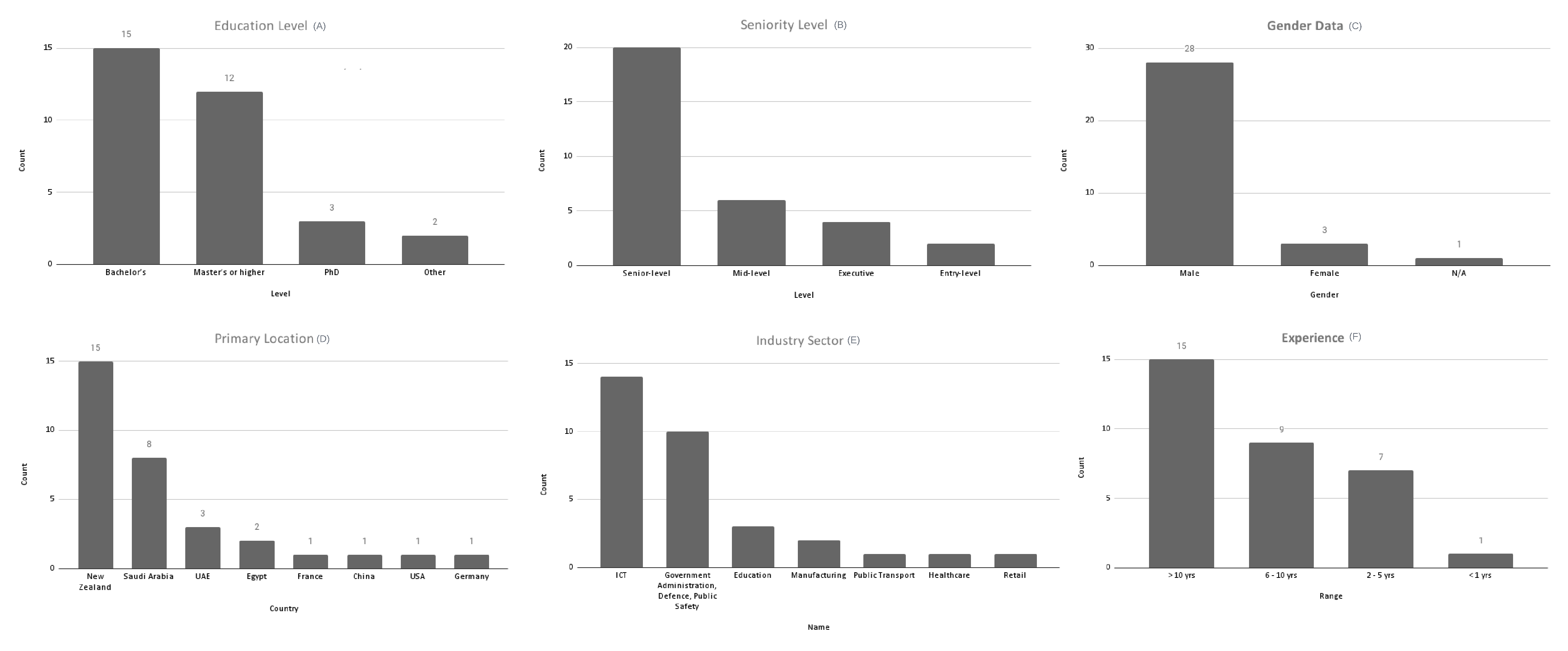}
    \caption{Survey Demographics}
    \label{fig:Demographics}
\end{sidewaysfigure}

For a full demographics overview (see Figure \ref{fig:Demographics}), below we present the highlights of the demographics:
\begin{enumerate}
\item Age: most participants are aged 35-44 (56\%), 25–34 (25\%), 45–54(16\%), and 55–64 (3\%).
\item Gender: 28 participants (87.5\%) are identified as male, and the remaining 12.5\% as female.
\item Experience: 47\% of the participants have more than 10 years of experience, followed by 9 participants (28\%) with 6-10 years of experience.
\item Education Level: Most participants have completed a Bachelor's degree in a computing-related discipline.
\item Agile Roles: the roles of the participants are varied, including developer (41\%), product owner, senior project manager, scrum master, and analysts.
\item Location: practitioners came from eight countries (New Zealand, Saudi Arabia, UAE, France, Egypt, China, USA, and Germany).
\item Industry Sectors: most participants primarily work in the ICT  sector (44\%). Other sectors (56\%) include government administration, defense, public safety, education, manufacturing, public transport, healthcare, and retail.
\item Seniority Levels: most participants hold senior-level positions (62\%).
\end{enumerate}

\textcolor{black}{Figure \ref{fig:Demographics} shows an overview of the demographic of the survey participants. We can see that the survey participants come from a diverse composition of roles, industries, and experience levels, offering a broad perspective on data management challenges in agile development. We can see that tmostparticipants are senior-level professionals (62\%), suggesting that the insights are informed by strategic and organizational viewpoints.  Almost half (47\%) of participants had over 10 years of professional expertise, representing mid-career professionals with significant industry experience. Most participants (94\%) held tertiary degrees in computing-related fields and occupied various agile roles. We note that participants represent different industrial sectors, with 44\% of participants from the ICT sector and the remainder from fields such as healthcare, education, public transport, and retail. This demographic composition highlights the study’s strength in capturing a broad spectrum of perspectives from practitioners across different industries. Although the survey is limited in its small sample size, the overrepresentation of participation from certain countries, and the predominance of specific roles, we made an effort to include a diverse range of industry sectors and seniority levels to overcome those limitations (discussed in more details in Section \ref{Full-ThreatstoValidity}).}

\clearpage

\subsubsection{Data Management Challenges and Solutions}
\textbf{Data Integration Challenges:}
The participants have flagged that \textit{managing data integration processes} is the most common challenge they face (62.50\%). \textit{Complexity of Integrating Real-Time Data} (44\%) is the least frequently mentioned challenge (see Figure \ref{fig:DICHRES}) and Table \ref{tab:AppDICH} in the Appendix. The participants highlighted that the integration process is challenging because of bugs and errors in the integration system used, data quality issues (such as data accuracy), data security issues (e.g., inaccessible data due to privacy concerns), and unclear integration requirements at the beginning of the project, which keep changing.

\begin{figure}[]
    \centering
    \includegraphics[width=\linewidth]{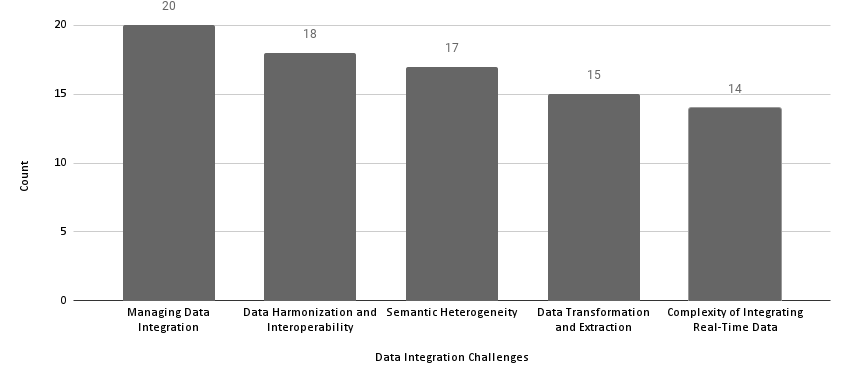}
   \caption{\textcolor{black}{Distribution of Data Integration Challenges Reported by Practitioners}}

    \label{fig:DICHRES}
\end{figure}

\textbf{Data Integration Solutions:}
Participants have identified the use of\\ \textit{communication-centric approaches} as the most frequently adopted solution for data integration challenges (56\%). \textit{Automated continuous testing} is the least frequently mentioned solution (34\%) (see Figure \ref{fig:DISOLRES} and Table \ref{tab:APPDISO} in the Appendix). Participants also employed solutions such as \textit{customized technical solutions}. For example, one participant noted that ``\textit{custom scripts were developed to automate the transformation of data into the required formats and structures before loading them into the target system}''. Another solution that the practitioners noted is \textit{standardization and governance} of data. Additionally, \textit{data segmentation and simplification} (i.e., dividing complex data integration into smaller ones) was also mentioned as a solution.

\begin{figure}[]
    \centering
    \includegraphics[width=\linewidth]{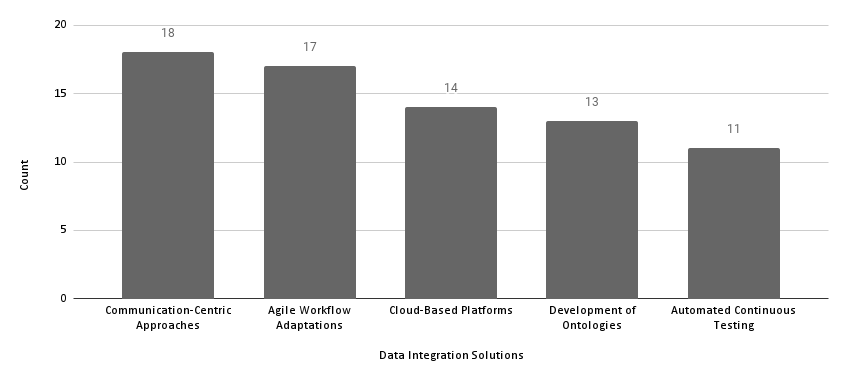}
        \caption{\textcolor{black}{Distribution of Data Integration Solutions Reported by Practitioners}}
    \label{fig:DISOLRES}
\end{figure}

\textbf{Data Collection Challenge:}
The participants have flagged that \textit{capturing diverse data} (59\%) and \textit{automation challenges} (59\%) are the most common data collection challenges. \textit{Comprehensive data collection} is the least frequently mentioned challenge (34\%) (see Figure \ref{fig:DCCH} and Table \ref{tab:AppDCCH} in the Appendix). The participants highlighted that the data collection is challenging because of the \textit{quality of the data being collected}, which in some cases might be incomplete or irrelevant. Other data collection challenges noted by participants include \textit{time Constraints} as the agile sprint time is insufficient to complete the tasks, and \textit{privacy issues} such as unavailable data to be collected due to ethical considerations.

\begin{figure}[]
    \centering
    \includegraphics[width=1\linewidth]{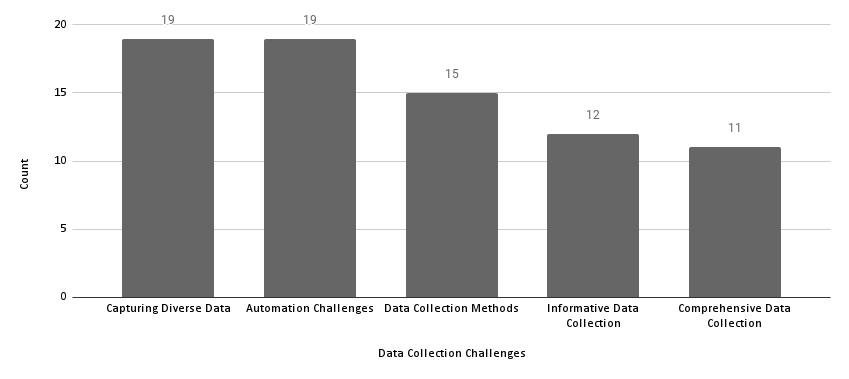}
     \caption{\textcolor{black}{Distribution of Data Collection Challenges Reported by Practitioners}}
    \label{fig:DCCH}
\end{figure}

\textbf{Data Collection Solutions:}
The participants have identified the use of \textit{user-centered design strategies} as the most frequently adopted solution for data collection challenges (59\%). \textit{Q-Rapids tool for valuable information} (3\%) is the least adopted solution (see Figure \ref{fig:DCSOL} and Table \ref{tab:APPDCSO} in the Appendix). The participants also employed solutions such as \textit{data transformation and standardization}. One participant noted transforming source data into a standardized format that can work across the source and target systems.

\begin{figure}[]
    \centering
    \includegraphics[width=1\linewidth]{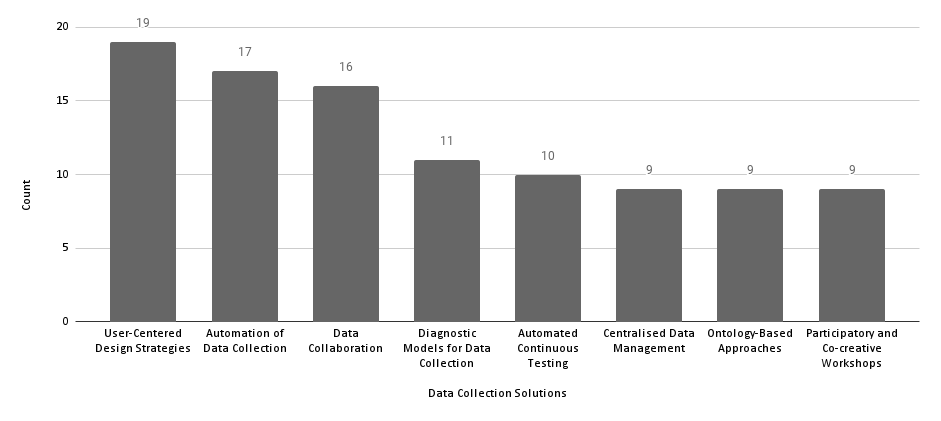}
     \caption{\textcolor{black}{Distribution of Data Collection Solutions Reported by Practitioners}}
    \label{fig:DCSOL}
\end{figure}

\textbf{Data Quality Challenge:}
The participants have flagged that \textit{ensuring data accuracy and consistency} (66\%) and \textit{completeness of data} (66\%) are the most common data quality challenges. \textit{Effective data quality management} (44\%) is the least frequently mentioned challenge (see Figure \ref{fig:DQCH} and Table \ref{tab:AppDQCH} in the Appendix). 


\begin{figure}[]
    \centering
    \includegraphics[width=1\linewidth]{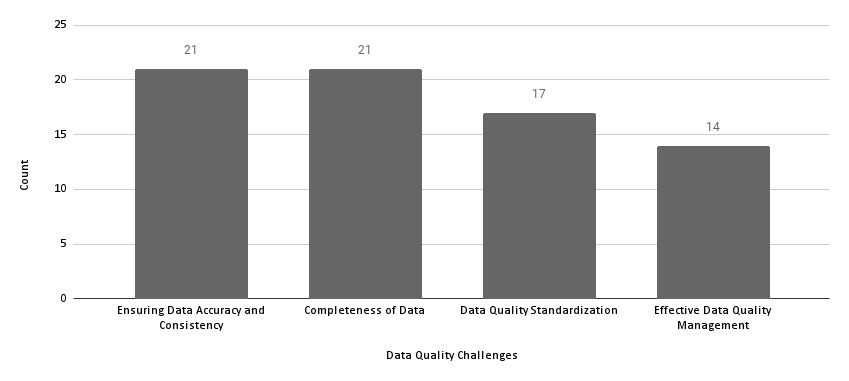}
      \caption{\textcolor{black}{Distribution of Data Quality Challenges Reported by Practitioners}}
    \label{fig:DQCH}
\end{figure}

\begin{figure}[]
    \centering
    \includegraphics[width=1\linewidth]{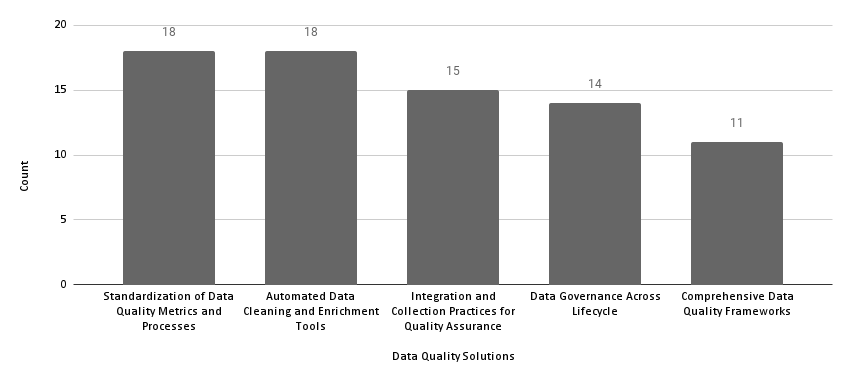}
    \caption{\textcolor{black}{Distribution of Data Quality Solutions Reported by Practitioners}}
    \label{fig:DQSOL}
\end{figure}

\textbf{Data Analysis Challenges:}
The participants have flagged \textit{that complex data sets} (59\%) and \textit{real-time analysis requirements} (59\%) are the most frequently reported challenges., followed by \textit{ensuring analytical accuracy} (47\%). \textit{Analyzing semantic heterogeneity} (25\%) and \textit{selection of appropriate analytical techniques} (25\%) are the least reported challenges (see Figure \ref{fig:DACH} and Table \ref{tab:AppDACH} in the Appendix). The participants highlighted that data analysis is challenging because of \textit{data integration and quality challenges}; for example, incomplete historical data due to integration difficulties or missing data can hinder the development of accurate predictive models. The participants mentioned that the \textit{data collection and preparation issues} may also impact data analysis. For example, it was mentioned that there is consistently a need ``\textit{to gather various data from various resources to model them together in order to get meaningful insights}''. Another data analysis challenge noted is \textit{unclear objective and scope issues} such as situations where the goals and boundaries of the data analysis tasks are not well-defined.

\begin{figure}[]
    \centering
    \includegraphics[width=1\linewidth]{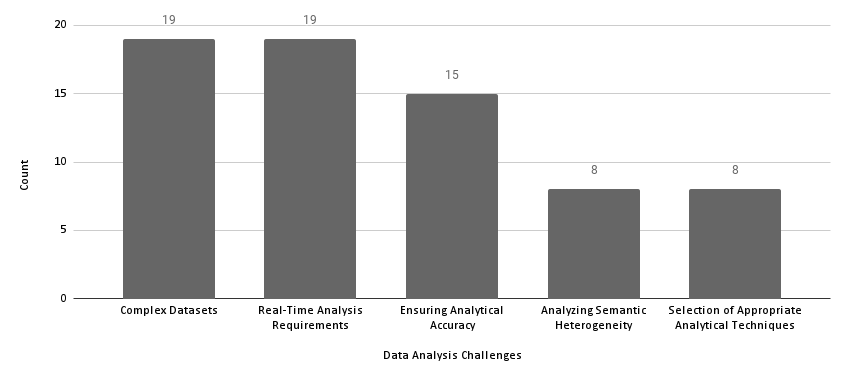}
      \caption{\textcolor{black}{Distribution of Data Analysis Challenges Reported by Practitioners}}
    \label{fig:DACH}
\end{figure}

\textbf{Data Analysis Solutions:}
Compared to the findings from our SLR, the participants have identified the use of \textit{advanced analytical tools and techniques} (50\%) as the most frequently adopted solution for data analysis challenges. \textit{Adoption of machine learning and AI for data analysis} (12.50\%) is the least reported solution (see Figure \ref{fig:DASOL} and Table \ref{tab:AppDASO} in the Appendix). The participants also highlighted that \textit{improve data integration, data quality, and collection methods} can facilitate the data analysis process.

\begin{figure}[]
    \centering
    \includegraphics[width=1\linewidth]{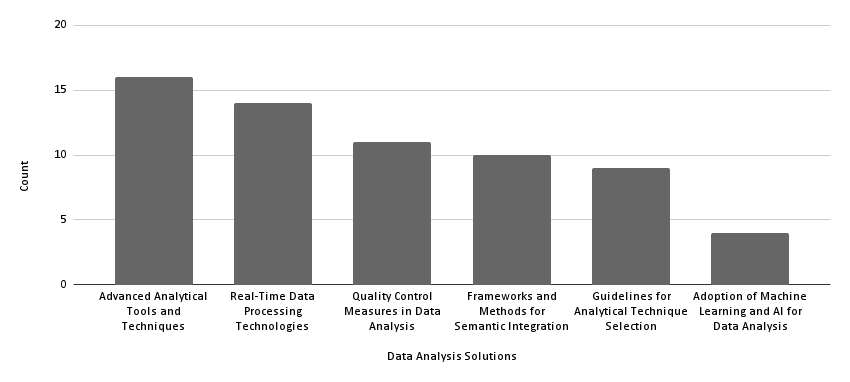}
      \caption{\textcolor{black}{Distribution of Data Analysis Solutions Reported by Practitioners}}
    \label{fig:DASOL}
\end{figure}

\subsection{\textcolor{black}{Alignment between the SLR and the Survey Findings}}

\textcolor{black}{The survey findings confirm the challenges and solutions identified in the SLR across all four data aspects we examined (data integration, collection, quality, and analysis). For data integration, 97\% (31 out of 32 participants) agreed with the SLR challenges. Additionally, 62.5\% of participants identified managing the data integration process as a significant challenge. They (97\%) supported solutions reported in the SLR, such as communication-centric methods (56\%). Similarly, for data collection, 91\% (29 out of 32 participants) agreed with the SLR data collection challenges, noting issues with capturing diverse data and automation (59\%). Furthermore, 87.5\% (29 participants) supported the proposed data collection solutions reported in the SLR, such as user-centered design strategies (59\%). For data quality, 94\% (30 participants) agreed with the data quality challenges reported in the SLR. They highlighted issues in ensuring accuracy and consistency (66\%), while 84\% (27 participants) agreed with the identified solutions, such as automated data cleaning and enrichment tools (56\%). With regard to data analysis, 81\% (26 participants) agreed with the data analysis challenges reported in the SLR. Specifically, 59\% of participants reported difficulties with large datasets and real-time requirements. Additionally, 78\% of the participants endorsed solutions such as advanced analytics tools (50\%).} 

\textcolor{black}{Overall, the majority of participants (91\%) approved of the data management challenges we identified from the SLR. A large percentage of the participants (87\%) also agreed with the proposed solutions noted in previous studies. These findings show a noteworthy alignment between the SLR and survey results, emphasizing the relevance of the identified challenges and the effectiveness of the proposed solutions.}

\subsection{Summary}

Our survey was conducted to complement the SLR findings with real-world insights from \textcolor{black}{industry agile practitioners who were significantly involved in data
management activities.} The survey involved 32 practitioners, predominantly senior-level professionals with over ten years of experience in agile software development. The survey confirmed many of the challenges identified in the SLR regarding data management aspects (RQ1). The results show that managing data integration processes, capturing diverse data, automation data collection challenges, ensuring data accuracy and consistency, completeness of data, complex data sets, and real-time analysis requirements are the most notable challenges faced by the participants. The participants adopted various solutions to address different data management challenges, which are mainly in line with the solutions reported in the SLR (RQ2). However, the survey results highlighted the need for training programs focusing on data management skills, which was not emphasized much in the previous studies. The survey findings show a greater need for real-time analytics and automation than indicated in the SLR. Detailed descriptions and examples of each challenge and solution are provided in the Appendix.

\section{Discussion}
\label{sec:discussion}
Our analysis of the literature in the SLR identified several challenges, including managing data integration processes, ensuring data accuracy and consistency, comprehensively collecting data, and managing complex data analysis. The practitioner survey (Section \ref{sec:survey}) confirms the majority of the data management challenges from the SLR (Section \ref{sec:slr}). 
 
\subsection{Comparison of the SLR and the Survey Findings}
The survey results also reflect data management challenges identified in the SLR. Most participants reported challenges with managing data integration processes, emphasizing the need for effective management of diverse data sources, which directly aligns with the SLR's findings on the complexities of managing data integration across various systems. Similarly, most survey respondents highlighted challenges with ensuring data accuracy, consistency, and completeness, which supports the SLR's emphasis on the importance of data quality for effective decision-making. 

The survey offered additional practical insights into data collection and analysis challenges. Most practitioners reported challenges regarding capturing diverse data during data collection and highlighted difficulties with automating data collection processes. These findings align with the SLR's emphasis on the need for comprehensive and standardized data collection methods. Regarding data analysis, most participants reported difficulties analyzing complex data sets and needing real-time analysis. This underscores the importance of advanced analytical tools and techniques, including communication-centric approaches and agile workflow adaptations, as solutions for complex data analysis.
The survey also showed a notable need for training programs that enhance data management skills within agile teams. This aspect was not given much emphasis in the studies in the SLR. Implementing real-time data analytics proved challenging for practitioners, who noted the need for better tools and training. The survey findings highlight a greater need for automation and real-time analytics than those mentioned in the SLR.

\subsection{Implications}
\label{sec:ImpRecom}
We explore the potential implications of our SLR and survey results. We discuss how these findings impact different agile process activities, including the impacted agile project roles and the potential strategies these roles can follow to mitigate possible challenges.

\noindent\textit{Requirements Gathering:} Inaccurate or incomplete data during requirements gathering can lead to poorly defined project goals and user stories, negatively impacting the entire development process and potentially leading to unmet stakeholder expectations. Product owners may struggle to define clear project goals. To mitigate this, they should ensure effective stakeholder communication and use comprehensive data collection methods. Business analysts may struggle to gather and document project requirements accurately. They should employ structured data-gathering techniques, use appropriate tools, and validate data with multiple sources. The project team may receive poorly defined requirements, leading to potential rework. To mitigate this, the team should collaborate closely with all stakeholders, thoroughly review requirements before implementation, and provide feedback to improve data collection methods.

\noindent\textit{Sprint Planning:} Comprehensive data collection is crucial during sprint planning for defining sprint goals and prioritizing tasks. Data management challenges ( e.g., inaccurate data collection) can lead to poorly defined sprint goals, affecting prioritization and sprint success and increasing the need for spike sprints (short, focused iterations designed to research or investigate a particular problem or uncertainty) \cite{al2020spikes}. Scrum masters may struggle to effectively facilitate the sprint planning session due to data inaccuracies. They should implement robust data validation processes, ensure clear communication of data management challenges, and facilitate collaborative problem-solving sessions. The product owners may face difficulties setting realistic sprint goals and prioritizing tasks. They should use data analytics tools to improve data accuracy, prioritize high-quality data sources, and regularly review data collection methods. The agile development team may end up working on poorly defined or prioritized tasks. Therefore, they may need extended workshops to understand and address the data management challenges, consuming valuable time and resources. To mitigate this, they should engage in continuous communication with the scrum master and product owner, participate in data validation, and ensure a clear understanding of sprint goals.

\noindent\textit{Daily Stand-ups:} Daily stand-ups are focused on identifying immediate obstacles and coordinating daily tasks. Data management challenges (e.g., completeness of data see table \ref{tab:AppDQCH}) can impede the ability to quickly identify blockers, and make informed decisions about the day's work. The project team may struggle to identify and communicate daily progress and blockers. They should use real-time data tracking tools, ensure daily stories data updates, and encourage open communication about data management challenges. Scrum masters may find it difficult to facilitate effective daily stand-ups. They should ensure accurate and timely data collection, address data management challenges promptly, and facilitate open discussions to resolve data-related challenges.

\noindent\textit{Retrospectives:} Retrospectives aims to reflect on the past iteration to understand what went well, what went wrong, and what could be improved \cite{dharmapal2016big,matthies2020playing}. Data management challenges (e.g., challenges in managing the integration process, such as integrating development project data into retrospective agendas) can hinder the agile development team's ability to analyze project performance data, derive actionable insights, and implement improvements. They should use data integration tools, ensure data accuracy and completeness, and facilitate collaborative analysis sessions. Moreover, scrum masters may struggle to guide the team in reflecting on past iterations and deriving actionable insights.  They should use structured retrospective techniques, ensure the availability of accurate data, and guide the team in data analysis. Product owners may face challenges in understanding the overall project performance and making informed decisions. They should use comprehensive data analytics tools, ensure regular data reviews, and involve stakeholders in the retrospective process.

\noindent\textit{Testing and Quality Assurance:} Data quality challenges necessitate better collaboration between testing teams and other cross-functional agile teams to ensure data integrity throughout the agile development lifecycle. The testing team should implement automated data quality checks, involve QA early in the data collection process, and ensure continuous collaboration with agile development teams. To ensure data accuracy, subject-matter experts should be involved early in the data collection and integration phases. 

As the industry continues to evolve towards more data-driven decision-making \cite{svensson2019unfulfilled,matthies2019towards}, the insights from the SLR and the survey can impact how practitioners approach data management in agile software development environments, leading to the adoption of new strategies that enhance agility, improve product quality, and facilitate better project outcomes. 

\subsection{Recommendations for Practitioners}
 We recommend developing comprehensive data management policies to address the data management challenges in support of agile methodologies used by practitioners. The data management policies require a collaborative effort among multiple stakeholders. Data governance team can lead the policy creation, ensuring alignment with regulatory requirements and organizational goals. Our SLR and survey show that including roles in the agile development team, such as legal advisors, can provide guidance on legal obligations and help the agile development team adhere to compliance with data privacy regulations \cite{barbala2023data}. The challenges that pertain to large volumes of data (e.g., ensuring data quality of large datasets) can handled by machine learning developers, who also help ensure that machine learning models are trained on accurate and representative datasets \cite{dautov2022towards}. Agile coaches assist teams in implementing data-driven decision-making policies, which help effectively utilize data to guide development processes \cite{batarseh2018predicting}. To develop comprehensive data management policies, we recommend that agile teams review the data management challenges and pick suitable solutions that suit their project and agile environment settings.
\section{Threats to Validity}
\label{Full-ThreatstoValidity}
In designing and conducting this study, we took careful steps to address various threats that could impact the validity of our findings \cite{ampatzoglou2020guidelines}, as discussed below:

\subsection{Internal Validity} 
To address possible search bias in our SLR, we first designed a \textit{comprehensive search strategy}. We employed a detailed search strategy with iterative pilot searches to refine the search terms (Section \ref{Subsec-SearchProcess}). This approach ensured that our search strings were broad enough to capture all relevant literature while being specific enough to exclude irrelevant studies to mitigate search string bias. 
We also defined a set of \textit{including and exclusion criteria} and applied them consistently during the study selection process (Section \ref{Subsec:Selection and Exclusion Criteria}) to reduce any selection bias.

To ensure a quality selection of articles, we conducted thorough quality checks to ensure each selected study's rigor, integrity, and relevance. Two authors evaluated each study using a predetermined quality assessment framework (see Section \ref{Sub-QA}), resolving any disagreements through discussion \cite{kitchenham2007guidelines}.

To minimize data extraction bias, we used \textit{a detailed data extraction procedure}. The first author conducted the initial data extraction, followed by verification through meetings with at least one other co-author. We adhered to thematic classification guidelines \cite{braun2006using} and used well-established methods for quantitative and qualitative data analysis \cite{kitchenham2007guidelines}.

\textcolor{black}{While conducting the SLR, we did not employ a literature snowballing process to screen for related articles cited in the SLR studies. This may have led to potentially missing relevant studies not being captured in our automated search. Although we initially reviewed references from the included studies, the additional actual studies identified were insufficient to justify their inclusion in our SLR. We adopted an informal snowballing approach (by checking the references of a subset of the nine most frequently appearing studies in our search) to cross-verify the results and ensure no significant studies were omitted. This approach involved examining references to enhance the robustness of our dataset.} 

In terms of the practitioner survey,  we iteratively designed the survey questions to ensure we designed a relevant and reliable questionnaire. This was done collaboratively between the researchers involved. We then piloted the questionnaire with three practitioners (with experience in data management) to ensure that the questions were relevant and understandable. This was done to ensure the survey was relevant and valid for the target audience. Additionally, we followed a detailed data analysis procedure with verification through meetings to minimize bias. The questionnaire development incorporated feedback from the pilot survey we ran with two agile practitioners who had experience dealing with data.

\subsection{External Validity}

The coverage provided by Scopus includes work from all relevant software engineering publication venues and enhances the generalizability of our findings \cite{kitchenham2007guidelines}.
We also followed a strict protocol-based SLR methodology outlined by Kitchenham and Charters \cite{kitchenham2007guidelines}. This facilitated the systematic extraction and review of data. While our findings may not be generalizable, they can be applied to various agile software development settings.

In the survey part of the study, we targeted a diverse group of practitioners from various industries and levels of experience in agile software development to ensure diversity and representation of the different roles. This diversity can help enhance our findings' generalizability to a broad range of agile software development environments. 

\textcolor{black}{The survey participants represented eight countries and worked on various agile development projects. However, we acknowledge the limitations imposed by the small sample size, the overrepresentation of participants from certain countries (e.g., New Zealand and Saudi Arabia), the limited representation from others (e.g., only one respondent each from France, Germany, and the USA), and the predominance of certain roles (e.g., developers and product owners) in our survey sample, which can restrict the generalizability of the findings. Still, we aimed to ensure that we have a diverse range of industry sectors and seniority levels represented in our sample.}

\subsection{Construct Validity}
To reduce the data extraction basis for both the SLR and the survey, we standardized data extraction forms to ensure uniformity. We refined our data extraction methods after several pilot tests. Two authors reviewed the extracted data process, resolving discrepancies through consensus. Each article that was extracted was then discussed between the researchers. We maintained a detailed description of the extraction process along with a replication package to ensure transparency and allow other researchers to verify our methods and findings \cite{fawzy_2024_10597818}.
The survey questionnaire was designed to accurately measure the intended constructs \cite{kitchenham2008personal,linaaker2015guidelines,robson2002real}. We considered convergent and divergent validity and integrated these into our survey design. We achieved this through the execution of a pilot survey, which helped to ensure the construct and content validity.

We also conducted inter-coder reliability checks and peer reviews to enhance the validity of the thematic analysis. An iterative process continuously improved the codes and themes, ensuring they were firmly grounded in the data and accurately reflected the participants' perspectives. Details of the thematic analysis process are explained in Section \ref{SureveyDataAnalysis}.

\subsection{Conclusion Validity}
To mitigate conclusion validity, we conducted a cross-validation process in which at least two authors analyzed the extracted data to cross-validate findings and interpretations, minimizing the likelihood of inaccurate conclusions (from both the SLR and the practitioner survey). An iterative review process involving team discussions and revisions based on collective feedback ensured the robustness of our conclusions.

\section{Conclusions}
\label{sec:conclusion}

In this study, we investigated data management challenges in agile software development and the proposed solutions through an SLR and a complementary survey with practitioners.
The SLR identified 45 studies, categorizing various aspects of data management such as data integration, quality, collection, and analysis. We then followed this analysis by surveying 32 practitioners with extensive experience with agile practices. This survey confirmed the main challenges we identified in our SLR and uncovered additional practical challenges that practitioners have reported. 
Our findings (from both the SLR and survey) reveal that managing data integration processes, capturing diverse data, automation data collection, ensuring data accuracy and consistency, completeness of data, complexity of data, and real-time analysis requirements are the most notable data management challenges faced by practitioners in agile development. Solutions, such as adopting ontology-based approaches and developing cloud-based platforms and automated tools, have shown promising results in mitigating these data management challenges. Furthermore, these data management challenges impact agile process activities and project roles, leading to poorly defined project goals and user stories. Product owners, business analysts, and scrum masters should use effective data management techniques to improve decision-making and overall project success.

Future studies should investigate advanced data integration techniques to harmonize and interoperate diverse and heterogeneous data sources. A promising research direction is using ontology-based, architecture-centric approaches and adopting AI techniques to resolve data management challenges. Furthermore, future studies should investigate systematic and automated data collection methodologies to enhance data accuracy and completeness throughout the software development life-cycle. This includes advanced automated quality assurance techniques combining real-time quality metrics and continuous monitoring tools. Such techniques and tools can help manage large volumes of structured and unstructured data and significantly improve agile teams' decision-making abilities. Future research should also rigorously evaluate the proposed solutions from both the SLR and practitioners' surveys to assess their effectiveness and applicability in real-world agile environments.

\section*{Conflicts of Interest}

The authors declared that they have no conflict of interest in the submission of this
manuscript.

\section*{Data Availability Statement}
A dataset of both the SLR and the practitioner survey is available online in our replication package \cite{fawzy_2024_10597818}.

\section*{Acknowledgments}
We would like to thank all practitioners who participated in our survey for their valuable time and input. Peng Liang is partially funded by NSFC under Grant No. 62172311 and the Major Science and Technology Project of Hubei Province under Grant No. 2024BAA008.

\appendix

\bibliographystyle{ieeetr}
\bibliography{references}

\clearpage
\section*{Appendix}
\label{sec:app}

\renewcommand{\thetable}{A\arabic{table}}
\setcounter{table}{0} 

\begin{table}[H]
\centering
\caption{Data Integration Challenges}
\label{tab:AppDICH}
\resizebox{\columnwidth}{!}{%
\begin{tabular}{ll}
\hline
\textbf{Challenges}                           & \textbf{Description and Example}                              \\ \hline
\multicolumn{1}{|l|}{Data Harmonization and Interoperability}  & \multicolumn{1}{l|}{\begin{tabular}[c]{@{}l@{}}Ensuring that data from different sources \\ can be understood and utilized together\\  by aligning common formats and schemas\\  without altering the original data content.\\  Example: Converting date formats from\\  MM/DD/YYYY in one system to \\ YYYY-MM-DD in another to unify \\ sales data analysis.\end{tabular}}                                   \\ \hline
\multicolumn{1}{|l|}{Semantic Heterogeneity}                   & \multicolumn{1}{l|}{\begin{tabular}[c]{@{}l@{}}Different meanings or interpretations \\ of the same data across various systems.\\  Example: Aligning differing patient \\ discharge status terms, like "Released" \\ vs. "Discharged," across multiple \\ hospital systems.\end{tabular}}                                                                                                                     \\ \hline
\multicolumn{1}{|l|}{Data Transformation and Extraction}       & \multicolumn{1}{l|}{\begin{tabular}[c]{@{}l@{}}Modifying data’s structure, format, \\ or content to meet the needs of a new\\  application or system, often involving\\  significant changes. \\ Example: Changing product sizes from \\ "S, M, L, XL" to detailed measurements\\  (e.g., "Small (S) - Chest: 34-36 inches")\\  and converting inventory data into JSON\\  for a modern system.\end{tabular}}  \\ \hline
\multicolumn{1}{|l|}{Managing Data Integration}                & \multicolumn{1}{l|}{\begin{tabular}[c]{@{}l@{}}Coordinating the process required\\  to merge data from disparate sources into \\ a unified system. Example: Integrating \\ customer data from online banking,\\  call centres, and branch visits into\\  a single data warehouse, requiring \\ a unified data model because\\  each of these departments uses \\ different systems and standards\end{tabular}} \\ \hline
\multicolumn{1}{|l|}{Complexity of Integrating Real-Time Data} & \multicolumn{1}{l|}{\begin{tabular}[c]{@{}l@{}}Merging and processing data that is\\  continuously generated from various\\  sources, requiring immediate \\ processing and integration for timely\\  decision-making . \\ Example: Integrating live stock market\\  feeds, transaction data from payment\\  systems, and customer feedback for \\ real-time analytics in a financial app.\end{tabular}}       \\ \hline
\end{tabular}}
\end{table}

\begin{table}[]
\centering
\caption{Data Collection Challenges}
\label{tab:AppDCCH}
\resizebox{\columnwidth}{!}{%
\begin{tabular}{ll}
\hline
\textbf{Challenges}                                            & \textbf{Description and Example}                                                    \\ \hline
\multicolumn{1}{|l|}{Capturing Diverse Data}                   & \multicolumn{1}{l|}{\begin{tabular}[c]{@{}l@{}}The challenge of collecting structured \\ and unstructured data from varied sources. \\ Example: Analyzing customer sentiment\\  requires collecting purchase data and\\  social media comments.\end{tabular}}                                                                                                                                                                                          \\ \hline
\multicolumn{1}{|l|}{Data Collection Method}                   & \multicolumn{1}{l|}{\begin{tabular}[c]{@{}l@{}}Difficulties arising from using different\\  data gathering tools or techniques. \\ Example: An agile development team \\ faces challenges with inconsistent data\\  collection from device sensors across\\  various smartphones. Differences in \\ sensor accuracy among smartphones\\  result in gaps in collecting user activity data\\  (what users are doing on their smartphones).\end{tabular}} \\ \hline
\multicolumn{1}{|l|}{Informative Data Collection}              & \multicolumn{1}{l|}{\begin{tabular}[c]{@{}l@{}}Ensuring that the collected data provides\\  the insights needed for decision-making. \\ Example: User activity data is collected,\\  but lacks details on specific user\\  experience issues.\end{tabular}}                                                                                                                                                                                            \\ \hline
\multicolumn{1}{|l|}{Comprehensive Data Collection}            & \multicolumn{1}{l|}{\begin{tabular}[c]{@{}l@{}}Systematically collecting data across \\ all project aspects. \\ Example: An e-commerce team might\\  miss out on user experience and \\ satisfaction data while focusing on \\ backend metrics.\end{tabular}}                                                                                                                                                                                          \\ \hline
\multicolumn{1}{|l|}{Automation Challenges in Data Collection} & \multicolumn{1}{l|}{\begin{tabular}[c]{@{}l@{}}Implementing automated systems for \\ data collection efficiently and accurately. \\ Example: An automated error logging system \\ incorrectly categorises user-generated errors,\\  complicating bug prioritisation.\end{tabular}}                                                                                                                                                                     \\ \hline
\end{tabular}}
\end{table}

\begin{table}[H]
\centering
\caption{Data Quality Challenges}
\label{tab:AppDQCH}
\resizebox{\columnwidth}{!}{%
\begin{tabular}{ll}
\hline
\textbf{Challenges}                                                                                                    & \textbf{Description and Example}                                                                                                                                                                                                                                                                                                                       \\ \hline
\multicolumn{1}{|l|}{\begin{tabular}[c]{@{}l@{}}Ensuring Data Accuracy and\\  Consistency Across Sources\end{tabular}} & \multicolumn{1}{l|}{\begin{tabular}[c]{@{}l@{}}Ensuring data from different sources is \\ accurate and presented in a uniform format. \\ Example: Integrating stock market data reveals\\  inconsistencies in trading volume reports from\\  different exchanges, requiring data\\  normalisation for accurate \\ application reporting.\end{tabular}} \\ \hline
\multicolumn{1}{|l|}{Completeness of Data}                                                                             & \multicolumn{1}{l|}{\begin{tabular}[c]{@{}l@{}}Ensuring every necessary data field is filled \\ without missing critical information. \\ Example: User stories for a new feature are\\  found to be missing acceptance \\ criteria during a sprint review, \\ leading to project delays and rework.\end{tabular}}                                      \\ \hline
\multicolumn{1}{|l|}{Effective Data Quality Management}                                                                & \multicolumn{1}{l|}{\begin{tabular}[c]{@{}l@{}}Implementing continuous practices to \\ uphold data integrity and address\\  issues promptly. \\ Example: Data corruption challenges \\ overwhelm automated correction efforts,\\  making manual reviews necessary and\\  impacting development timelines.\end{tabular}}                                \\ \hline
\multicolumn{1}{|l|}{Data Quality Standardization}                                                                     & \multicolumn{1}{l|}{\begin{tabular}[c]{@{}l@{}}Establishing and applying uniform measures\\  for data quality assessment \\ across varied projects. \\ Example: A team faces difficulty creating\\  data quality metrics that are relevant and\\  applicable to all projects, hindering\\  company-wide quality assessments.\end{tabular}}             \\ \hline
\end{tabular}}
\end{table}

\begin{table}[]
\centering
\caption{Data Analysis Challenges}
\label{tab:AppDACH}
\resizebox{\columnwidth}{!}{%
\begin{tabular}{ll}
\hline
\textbf{Challenges}                                  & \textbf{Description and Example}                                                                                                                                                                                                                                                                                                                     \\ \hline
\multicolumn{1}{|l|}{Complex Data Sets}                              & \multicolumn{1}{l|}{\begin{tabular}[c]{@{}l@{}}Challenges posed by analysing large \\ or complex datasets. \\ Example: A team refines vast user\\  data over many sprints for to build \\ a machine learning project.\end{tabular}}                                                                                                                  \\ \hline
\multicolumn{1}{|l|}{Real-Time Analysis Requirements}                & \multicolumn{1}{l|}{\begin{tabular}[c]{@{}l@{}}The challenge of analysing data\\  as it is generated, without delay, \\ to support immediate decision-making \\ or operational actions. \\ Example: Developing a real-time user\\  engagement tracker, frequently revisiting \\ their approach to meet analysis\\  speed requirements.\end{tabular}} \\ \hline
\multicolumn{1}{|l|}{Analyzing Semantic Heterogeneity Data}          & \multicolumn{1}{l|}{\begin{tabular}[c]{@{}l@{}}Difficulties with data with varying meanings\\  across sources. \\ Example: A healthcare team struggles \\ to align medical procedure \\ terms across datasets.\end{tabular}}                                                                                                                         \\ \hline
\multicolumn{1}{|l|}{Ensuring Analytical Accuracy}                   & \multicolumn{1}{l|}{\begin{tabular}[c]{@{}l@{}}Keeping data analysis precise and \\ reliable to minimize errors. \\ Example: A financial project team \\ uses pair programming to boost analysis\\  accuracy, adding unforeseen tasks\\  to their agile process.\end{tabular}}                                                                       \\ \hline
\multicolumn{1}{|l|}{\begin{tabular}[c]{@{}l@{}}Selection of Appropriate\\ Analytical Techniques\end{tabular}} & \multicolumn{1}{l|}{\begin{tabular}[c]{@{}l@{}}Choosing suitable analysis methods\\  based on data characteristics. \\ Example: Debating over using \\ statistical models or machine \\ learning for developing \\ a fraud detection system\end{tabular}}                                                                                            \\ \hline
\end{tabular}}
\end{table}


\begin{table}[H]
\centering
\caption{Data Integration Solutions}
\label{tab:APPDISO}
\resizebox{\columnwidth}{!}{%
\begin{tabular}{|l|l|}
\hline
\textbf{Solutions}               & \textbf{Description and Example}                                                                                                           \\ \hline
Development of Ontologies        & \begin{tabular}[c]{@{}l@{}}Using tools or guides to create a unified vocabulary\\  for data across systems . \\ Example: Employing an ontology editor\\  like Protégé tool to build ontologies \\ that standardize company-wide data terms.\end{tabular}                                                                        \\ \hline
Cloud-Based Platforms            & \begin{tabular}[c]{@{}l@{}}Using cloud services for efficient \\ data management and integration. \\ Example: Utilizing AWS for scalable \\ data ingestion and transformation.\end{tabular}                                                                                                                                     \\ \hline
Communication-Centric Approaches & \begin{tabular}[c]{@{}l@{}}Applying communication theory \\ to harmonise data meanings across teams. \\ Example: Holding regular team meetings \\ to ensure consistent data interpretation \\ among stakeholders.\end{tabular}                                                                                                  \\ \hline
Automated Continuous Testing     & \begin{tabular}[c]{@{}l@{}}Employing automated testing frameworks. \\ Example: Using ( Jenkins, Travis CI, \\ or GitHub Actions) to regularly check data quality\\  and integrity as part of the development process.\end{tabular}                                                                                              \\ \hline
Agile Workflow Adaptations       & \begin{tabular}[c]{@{}l@{}}Adjusting development practices to improve \\ data integration outcomes. \\ Example: integrating data quality checks\\  into daily stand-ups, adjusting sprint reviews \\ to include data integration retrospectives, \\ or adopting pair programming for \\ complex integration tasks.\end{tabular} \\ \hline
\end{tabular}}
\end{table}

\begin{table}[H]
\centering
\caption{Data Collection Solutions}
\label{tab:APPDCSO}
\resizebox{\columnwidth}{!}{%
\begin{tabular}{|l|l|}
\hline
\textbf{Solutions}                                                                                     & \textbf{Description and Example}                                                                                                                                                                                    \\ \hline
User-Centered Design Strategies                                                                        & \begin{tabular}[c]{@{}l@{}}Tailoring data collection methods\\  to be intuitive and user-focused. \\ Example: Designing surveys\\  and interaction trackers based \\ on user behavior studies.\end{tabular}         \\ \hline
\begin{tabular}[c]{@{}l@{}}Automated Continuous Testing \\ and Quality Metrics Dashboards\end{tabular} & \begin{tabular}[c]{@{}l@{}}Monitoring data quality through \\ automated tools and real-time dashboards. \\ Example: Implementing continuous \\ testing to quickly spot and rectify data issues.\end{tabular}        \\ \hline
\begin{tabular}[c]{@{}l@{}}Automation of Data Collection\\  and Visualization Toolchains\end{tabular}  & \begin{tabular}[c]{@{}l@{}}Using automated systems for efficient \\ data gathering and visualization. \\ Example: Streamlining data workflows\\  to enhance collection accuracy \\ and analysis speed.\end{tabular} \\ \hline
Centralised Data Management Systems                                                                    & \begin{tabular}[c]{@{}l@{}}Centralizing data handling to improve \\ consistency and data access. \\ Example: Creating a unified \\ data platform for all stages\\  from collection to analysis.\end{tabular}        \\ \hline
\begin{tabular}[c]{@{}l@{}}Diagnostic Models for Data Collection\\  and Sharing Practices\end{tabular} & \begin{tabular}[c]{@{}l@{}}Assessing and refining data practices \\ to improve quality and cooperation. \\ Example: Developing models to \\ optimise data gathering and \\ sharing workflows.\end{tabular}          \\ \hline
\begin{tabular}[c]{@{}l@{}}Data-Driven Systems Engineering \\ (DDSE) Methodologies\end{tabular}        & \begin{tabular}[c]{@{}l@{}}Integrating data analysis into \\ system engineering for informed decisions. \\ Example: Applying DDSE for \\ collaborative and data-informed\\ project development.\end{tabular}        \\ \hline
\begin{tabular}[c]{@{}l@{}}Legal Advisor for \\ Data Sensitivity Assessment\end{tabular}               & \begin{tabular}[c]{@{}l@{}}Ensuring data collection complies with \\ data protection laws. \\ Example: Consulting legal experts to navigate\\  privacy regulations in data practices.\end{tabular}                  \\ \hline
\begin{tabular}[c]{@{}l@{}}Development of Ontology-Based\\  Approaches\end{tabular}                    & \begin{tabular}[c]{@{}l@{}}Enhancing data usefulness through \\ structured data categorisation. \\ Example: Using ontologies to \\ standardize data terms for \\ better interoperability.\end{tabular}              \\ \hline
\begin{tabular}[c]{@{}l@{}}Q-Rapids Tool for Valuable \\ Information Acquisition\end{tabular}          & \begin{tabular}[c]{@{}l@{}}Incorporating tools designed \\ for real-time data analysis in development. \\ Example: Leveraging Q-Rapids for focused \\ data collection in software projects.\end{tabular}            \\ \hline
\begin{tabular}[c]{@{}l@{}}Agile Methodology and \\ Data Analyst Collaboration\end{tabular}            & \begin{tabular}[c]{@{}l@{}}Jointly developing data strategies with \\ agile teams and data analysts. \\ Example: Collaborating on adaptable \\ data collection methods suited to project needs.\end{tabular}        \\ \hline
\begin{tabular}[c]{@{}l@{}}Participatory and \\ Co-creative Workshops\end{tabular}                     & \begin{tabular}[c]{@{}l@{}}Engage stakeholders in refining data collection \\ and integration. Example: Hosting workshops \\ to gather diverse inputs on \\ improving data practices.\end{tabular}                  \\ \hline
\end{tabular}}
\end{table}

\begin{table}[H]
\centering
\caption{Data Quality Solutions}
\label{tab:AppDQSO}
\resizebox{\columnwidth}{!}{%
\begin{tabular}{|l|l|}
\hline
\textbf{Solutions}                                                                                    & \textbf{Description and Example}                                                                                                                                                                                                                                                                                          \\ \hline
\begin{tabular}[c]{@{}l@{}}Comprehensive Data Quality\\  Frameworks\end{tabular}                      & \begin{tabular}[c]{@{}l@{}}Using established guidelines for ongoing \\ data quality improvement. \\ Example: Applying Total Data Quality\\  Management (TDQM) principles to \\ ensure data quality across the organisation.\end{tabular}                                                                                  \\ \hline
\begin{tabular}[c]{@{}l@{}}Automated Data Cleaning\\  and Enrichment Tools\end{tabular}               & \begin{tabular}[c]{@{}l@{}}Utilizing software to automatically correct\\  and enhance data. \\ Example: Deploying Tools like Talend \\ or Trifacta for real-time data cleaning \\ and standardisation.\end{tabular}                                                                                                       \\ \hline
Data Governance Across Lifecycle                                                                      & \begin{tabular}[c]{@{}l@{}}Creating rules and roles for \\ data management from creation to deletion.\\  Example: Establishing a governance \\ committee to oversee \\ data security and privacy.\end{tabular}                                                                                                            \\ \hline
\begin{tabular}[c]{@{}l@{}}Standardization of Data Quality\\  Metrics and Processes\end{tabular}      & \begin{tabular}[c]{@{}l@{}}Setting uniform data quality standards\\ and tracking mechanisms. \\ Example: Developing a set of \\ data quality KPIs (e.g., accuracy, \\ completeness, consistency, timeliness)\\  for data accuracy and consistency, \\ monitored using tools like\\  data quality scorecards.\end{tabular} \\ \hline
\begin{tabular}[c]{@{}l@{}}Integration and Collection Practices\\  for Quality Assurance\end{tabular} & \begin{tabular}[c]{@{}l@{}}Embedding quality checks into \\ data collection and integration workflows.\\  Example: Incorporating validation s\\ teps in ETL processes to ensure \\ data integrity from the start.\end{tabular}                                                                                            \\ \hline
\begin{tabular}[c]{@{}l@{}}Integration of Advanced \\ Analytic Platforms\end{tabular}                 & \begin{tabular}[c]{@{}l@{}}Adopting high-tech platforms for \\ enhanced data management and quality.\\ Example: Using Apache Kafka for seamless\\  real-time data integration and quality analysis.\end{tabular}                                                                                                          \\ \hline
\end{tabular}}
\end{table}

\begin{table}[H]
\centering
\caption{Data Analysis Solutions}
\label{tab:AppDASO}
\resizebox{\columnwidth}{!}{%
\begin{tabular}{|l|l|}
\hline
\textbf{Solutions}                                                                               & \textbf{Description and Example}                                                                                                                                                                                                                                           \\ \hline
\begin{tabular}[c]{@{}l@{}}Advanced Analytical\\  Tools and Techniques\end{tabular}              & \begin{tabular}[c]{@{}l@{}}Employing Utilizing high-level software \\ for in-depth analysis of complex datasets.\\  Example: Using R and Python for advanced\\  data exploration and Tableau for visualisation.\end{tabular}                                               \\ \hline
\begin{tabular}[c]{@{}l@{}}Real-Time Data\\  Processing Technologies\end{tabular}                & \begin{tabular}[c]{@{}l@{}}Using technologies that enable immediate \\ analysis of live data streams. \\ Example: Utilizing big data frameworks\\  for streaming and on-the-fly data processing\\  enables organizations to act on\\  insights without delay.\end{tabular} \\ \hline
\begin{tabular}[c]{@{}l@{}}Frameworks and Methods\\  for Semantic Integration\end{tabular}       & \begin{tabular}[c]{@{}l@{}}Applying organised methods to harmonise\\  data meanings across sources. \\ Example: Utilizing RDF \\ (Resource Description Framework) \\ and SPARQL queries with ontologies\\  for consistent data interpretation.\end{tabular}                \\ \hline
\begin{tabular}[c]{@{}l@{}}Quality Control Measures\\  in Data Analysis\end{tabular}             & \begin{tabular}[c]{@{}l@{}}Establishing checks to maintain\\  the precision of data analysis outcomes.\\  Example: Integrating validation steps\\  and cross-validation in analysis\\  workflows to ensure accuracy.\end{tabular}                                          \\ \hline
\begin{tabular}[c]{@{}l@{}}Guidelines for Analytical\\  Technique Selection\end{tabular}         & \begin{tabular}[c]{@{}l@{}}Creating criteria to assist in selecting\\  suitable data analysis methods. \\ Example: Developing a decision\\  matrix to guide the choice between\\  statistical and machine learning \\ techniques based on data specifics.\end{tabular}     \\ \hline
\begin{tabular}[c]{@{}l@{}}Adoption of Machine Learning\\  and AI for Data Analysis\end{tabular} & \begin{tabular}[c]{@{}l@{}}Utilizing AI and machine learning\\  to enhance pattern recognition and \\ prediction in data. \\ Example: Leveraging AI and libraries\\  such as TensorFlow and PyTorch\\  for automated insights \\ and predictive analysis.\end{tabular}     \\ \hline
\end{tabular}}
\end{table}

\begin{table}[]
\caption{Data Types as Discussed in the Studies}
\label{tab:DataTypes}
\resizebox{\columnwidth}{!}{%
\begin{tabular}{llllll}
\toprule
Study                                          & Process Data & Project Data & Product Data & Operational data & Bussiness Data \\
\midrule
\cite{chen2016agile}              & X            &              & X            &                  &                \\
\cite{pater2018advancing}         &              &              &              & X                & X              \\
\cite{rosenkranz2017supporting}   &              &              &              &                  & X              \\
\cite{canedo2022guidelines}       & X            & X            &              &                  &                \\
\cite{harper2014agile}            &              &              & X            & X                &                \\
\cite{grunewald2021cloud}         &              &              &              & X                &                \\
\cite{svensson2019unfulfilled}    &              &              &              & X                & X              \\
\cite{dautov2022towards}          &              &              &              & X                &                \\
\cite{fabijan2016lack}            & X            & X            &              &                  &                \\
\cite{dursun2014workflow}         &              &              &              & X                &                \\
\cite{dharmapal2016big}           &              &              &              &                  & X              \\
\cite{das2015towards}             &              &              &              &                  & X              \\
\cite{min2010practices}           &              &              &              & X                &                \\
\cite{vestues2022agile}           &              &              &              &                  & X              \\
\cite{barbala2023data}            &              &              &              & X                & X              \\
\cite{abdallah2022towards}        &              &              &              & X                &                \\
\cite{olsson2018challenges}       &              &              &              & X                &                \\
\cite{rix2016agile}               &              &              &              & X                &                \\
\cite{spengler2020enabling}       &              &              &              & X                &                \\
\cite{kannan2017rapid}            &              &              &              & X                &                \\
\cite{schuttler2021journey}       &              & X            & X            &                  &                \\
\cite{hofer2020new}               & X            &              & X            &                  &                \\
\cite{huang2021leveraging}        &              &              &              &                  & X              \\
\cite{vogt2023implementing}       &              &              &              & X                & X              \\
\cite{upender2005staying}         & X            & X            & X            &                  &                \\
\cite{bosch2019towards}           &              &              &              & X                & X              \\
\cite{ambler2008gets}             & X            & X            &              &                  &                \\
\cite{harriman2004emergent}       & X            &              & X            &                  &                \\
\cite{chung2006bridging}          &              &              & X            &                  &                \\
\cite{dos2021ontology}            & X            &              & X            &                  &                \\
\cite{matthies2019towards}        & X            &              &              &                  & X              \\
\cite{franch2019quality}          & X            &              & X            &                  &                \\
\cite{chhillar2019act}            & X            &              & X            & X                &                \\
\cite{jenness2018lsst}            & X            & X            &              &                  &                \\
\cite{batarseh2018predicting}     & X            & X            &              &                  &                \\
\cite{alsaadi2022data}            & X            & X            &              &                  &                \\
\cite{kaur2020dialogue}           & X            & X            &              &                  &                \\
\cite{matthies2020playing}        & X            & X            &              &                  &                \\
\cite{martinez2019continuously}   & X            & X            &              &                  &                \\
\cite{lin2018towards}             & X            & X            &              &                  &                \\
\cite{little2004adaptive}         &              & X            &              &                  &                \\
\cite{barcellos2020towards}       & X            & X            &              &                  &                \\
\cite{lehtonen2017visualizations} & X            & X            &              &                  &                \\
\cite{hamer2023students}          & X            & X            &              &                  &                \\
\cite{fagarasan2023integrating}   & X            & X            &              &                  &     \\
\bottomrule
\end{tabular}}
\end{table}

\end{document}